\documentclass[usenatbib]{mn2e}
\usepackage{epsfig}
\usepackage[pass]{geometry}
\usepackage{times}

\title[SN2011ja Evolution]{Early Dust Formation and a Massive Progenitor for SN 2011ja?}
\author[Andrews et al.]{J.E. Andrews$^1$\thanks{Email: jandrews@as.arizona.edu}, 
Kelsie M. Krafton$^2$, Geoffrey C. Clayton$^2$, E. Montiel$^2$, R. Wesson$^3$,
\newauthor
Ben E.K. Sugerman$^4$, M.J. Barlow$^5$, M. Matsuura$^6$, \& H. Drass$^7$\\
$^1$Steward Observatory, University of Arizona, 933 North Cherry Avenue, Tucson, AZ 85721, USA\\
 $^2$Department of Physics and Astronomy, Louisiana State University, 202 Nicholson Hall, Baton Rouge, LA 70803, USA\\
 $^3$European Southern Observatory, Alonso de C´ordova 3107,19001 Casilla, Santiago, Chile\\
 $^4$Department of Physics and Astronomy, Goucher College, 1021 Dulaney Valley Rd., Baltimore, MD 21204\\
 $^5$Department of Physics and Astronomy, University College London, Gower Street, London WC1E 6BT\\
 $^6$School of Physics and Astronomy, Cardiff University,Cardiff CF24 3AA, UK\\
 $^7$Astronomisches Institut, Ruhr-Universit\"at Bochum, Universit\"atsstra\ss{}e 150, 44780 Bochum, Germany}

\begin{document}
\date{Accepted 0000, Received 0000, in original form 0000}
\pagerange{\pageref{firstpage}--\pageref{lastpage}} \pubyear{2015}
\def\arcdeg{\degr}
\maketitle
\label{firstpage}

\begin{abstract}
SN 2011ja was a bright (I = -18.3) Type II supernova occurring in the nearby edge on spiral galaxy NGC 4945. Flat-topped and multi-peaked  H$\alpha$ and H$\beta$ spectral emission lines appear between 64 - 84 days post-explosion, indicating interaction with a disc-like circumstellar medium inclined 30-45 degrees from edge-on.  After day 84 an increase in the H- and K-band flux along with heavy attenuation of the red wing of the emission lines are strong indications of early dust formation, likely located in the cool dense shell created between the forward shock of the SN ejecta and the reverse shock created as the ejecta plows into the existing CSM.  Radiative transfer modeling reveals both $\approx$ 1.5 $\times$ 10$^{-4}$ M$_{\sun}$ of  pre-existing dust located $\sim$ 10$^{16.7}$ cm away and $\approx$ 5 $\times$ 10$^{-5}$ M$_{\sun}$ of newly formed dust.  Spectral observations after 1.5 years reveal the possibility that the fading SN is located within a young (3-6 Myr) massive stellar cluster, which when combined with tentative $^{56}$Ni mass estimates of 0.2 M$_{\sun}$ may indicate a massive ($\geq$ 25 M$_{\sun}$) progenitor for SN 2011ja. 
\end{abstract}


\section{Introduction}

Type IIP, the most common type of core collapse supernovae (CCSNe), have broad ($\sim$10$^4$ km s$^{-1}$) hydrogen emission lines along with a near constant ``plateau" of optical luminosity throughout the first $\sim$100 days.  The widely accepted progenitors of Type IIP SNe, red supergiants (RSGs), have masses ranging between $\sim$9-25 M$_{\sun}$ and mass loss rates of $\sim$10$^{-6}$ to 10$^{-4}$ M$_{\sun}$ yr$^{-1}$ \citep{2006ApJ...641.1029C, 2011A&A...526A.156M}. Type IIn SNe also show broad hydrogen emission, but, in addition, they show narrow ($\sim$ 100 km s$^{-1}$) hydrogen emission due to ionization of the surrounding, dense pre-existing circumstellar material (CSM).  These SNe likely have more massive progenitors such as Luminous Blue Variables (LBV), or Wolf-Rayet (WR) stars that can have mass loss rates which are orders of magnitude larger, 10$^{-5}$ to 10$^{-2}$ M$_{\sun}$ yr$^{-1}$ \citep{2012ApJ...744...10K, 2011ApJ...732...63S, 2014MNRAS.438.1191S}.  Whether Type IIP or Type IIn, these massive star progenitors can undergo periods of dramatic mass loss prior to explosion which has a direct impact on the SN evolution. Over the past decade we have seen numerous observational signatures of the SN ejecta interacting with previously shed layers, anywhere from hours to years after explosion \citep[for example]{2000ApJ...536..239L, 2010ApJ...715..541A, 2014ApJ...797..118F, 2015MNRAS.449.1876S,2015arXiv150406668M}. By observing the SN-CSM interaction we can get a better understanding of pre-supernova mass loss, pathways of dust production, and the link between progenitor and SN type.

\begin{figure*} 
   \centering
   \includegraphics[width=7in]{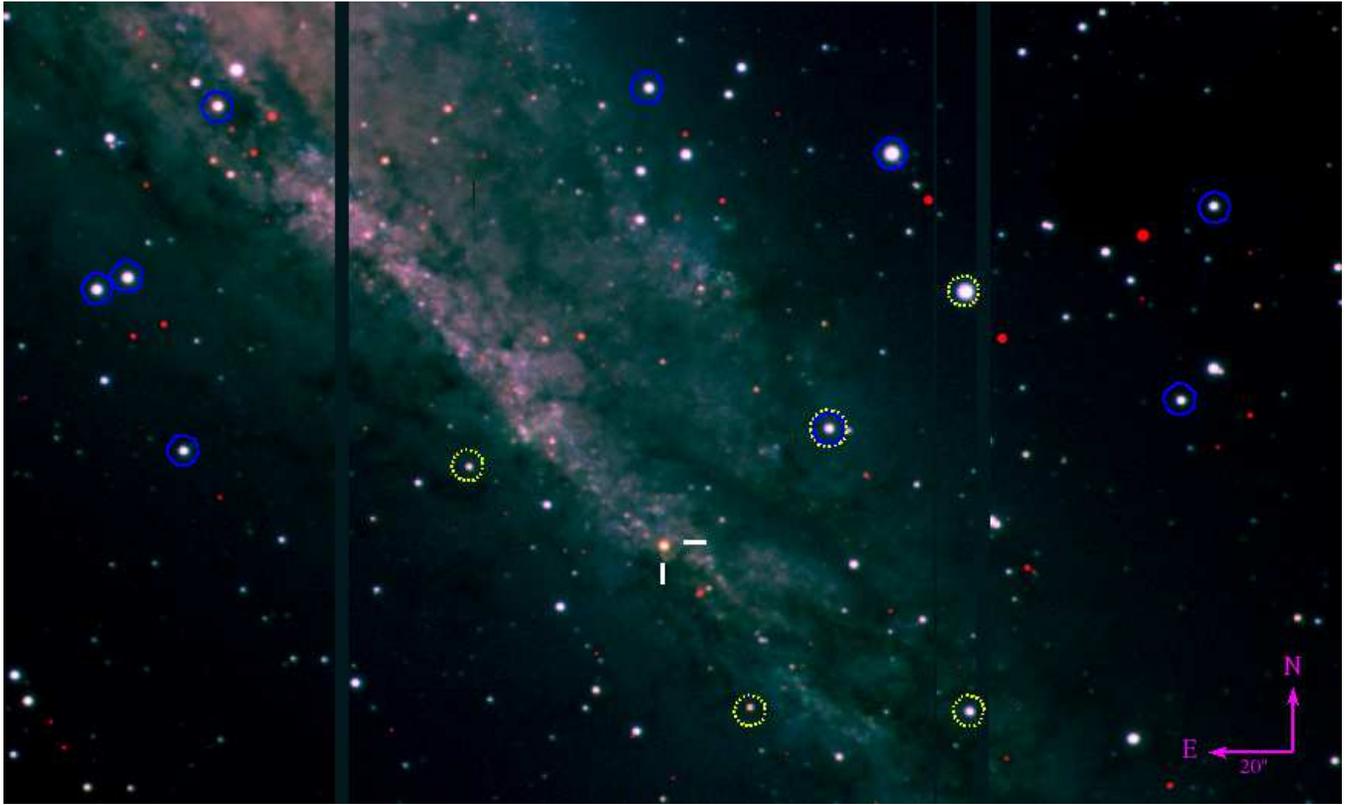} 
   \caption{Color composite Gemini/GMOS image of SN 2011ja (indicated in white).  Optical standards (Table 3) are indicated by blue solid circles, NIR standards (Table 4) by dashed yellow circles.  }
   \label{fig:example}
\end{figure*}

Due to their short lifetimes and ability to return material back to the ISM quickly, CCSNe are the likely culprits for the dust production in dusty high-z galaxies \citep{2011A&A...528A..14G,2014ApJ...788L..30D}. This is a double-edged sword, as they are also efficient destroyers of dust, at least in present day galaxies \citep{2015ApJ...799..158T}.  Although \citet{2015ApJ...803....7S} propose effective SN dust destruction, but only within a subset of the appropriate assumed parameters.  Many recent studies of nearby CCSNe have searched for signatures of dust formation and estimated the dust masses. The results of these studies indicate small amounts of newly-formed dust, 10$^{-2}$ - 10$^{-4}$ M$_{\sun}$ \cite[for example]{2003MNRAS.338..939E,2006Sci...313..196S,2007ApJ...665..608M,2009ApJ...704..306K,2010ApJ...715..541A}, much less than the 0.1-1  M$_{\sun}$ needed to account for the excess of dust seen in the early Universe. Within the past few years, far-IR studies have revealed a few sources which may hold promise to unlocking the dust mystery.  Herschel observations of three nearby young supernova remnants indicate the presence of 0.1 M$_{\sun}$ of cool dust in Cas A \citep{2010A&A...518L.138B}, 0.4 -- 0.7 M$_{\sun}$ of cool  dust in the ejecta of SN 1987A \citep{2011Sci...333.1258M, 2015ApJ...800...50M} and 0.18-0.27 M$_{\sun}$ of cool dust in the Crab Nebula \citep{2012ApJ...760...96G, 2015ApJ...801..141O}. Recent SOFIA observations of a supernova remnant at the center of the Milky Way revealed 0.02 M$_{\sun}$ of warm (100 K) dust, which also has seemed to have survived the passing of the reverse shock \citep{2015arXiv150307173L}. This strongly suggests that almost all of the dust in SNe is formed after $>$ 1000 days \citep{2015MNRAS.446.2089W, 2015arXiv150900858B}. While reservoirs of cold dust may help solve this problem, there is a growing body of evidence that grain growth may be occurring in the ISM enriched by these early SNe and could be the main site for dust production at high-z \citep{2015A&A...577A..80M}. However, a plausible mechanism for growing refractory grains in the ISM has yet to be identified.

Observational signatures of dust formation in CCSNe manifest themselves in several different forms.  The optical luminosity will decrease while almost simultaneously the NIR will increase as the dust grains absorb the shorter wavelength light and re-emit it in the IR.  The grain formation will also alter the optical spectra, creating asymmetric and blue-shifted lines as the dust grains attenuate the red, receding side of the ejecta preferentially.  This observational evidence of dust formation has now been seen in numerous SNe, including the nearby and well-studied SN 1987A \citep{1989LNP...350..164L,1993ApJS...88..477W}, SN 2003gd \citep{2006Sci...313..196S,2007ApJ...665..608M}, SN 2004et \citep{2006MNRAS.372.1315S,2009ApJ...704..306K} and many others.  While it was initially believed that the dust grains could only condense 300-600 days after explosion, when the ejecta had expanded and cooled, there have been more and more confirmed cases of dust forming much earlier, within $\sim$100 days of explosion.  This can occur due to shock interaction with nearby CSM creating an area between the forward and reverse shocks with temperatures and densities appropriate for grain growth, this area is known as the cool dense shell (CDS). For example SN 1998S showed dust formation signatures between days 140 - 268 \citep{2000ApJ...536..239L}, and SN 2005ip appears to have formed dust both in the CDS between day 75-150 and then again in the ejecta after day 750 \citep{2009ApJ...695.1334S,2009ApJ...691..650F}. The bright IIn SN 2010jl  shows continuous dust formation between 40 and 240 days \citep{2014Natur.511..326G}. Although not classified as Type IIn, the Type Ib/c SN 2006jc also formed dust via CSM interaction between 50 - 75 days post-explosion \citep{2008MNRAS.389..141M,2008ApJ...680..568S} and the Type IIP SN 2007od formed dust sometime between day 120 - 230 through the same mechanism \citep{2010ApJ...715..541A}.  As we will present below, with more and more long-term monitoring of CCSNe, there seems to be evidence of non-Type-IIn SNe exhibiting signs of CSM interaction, even months after explosion.  This not only allows a separate channel for dust formation in CCSNe, but can also reveal important properties of SN evolution.

SN 2011ja was discovered in NGC 4945, an edge-on spiral (Figure 1) located at a distance of 3.36 $\pm$ 0.09 Mpc \citep{2005ApJ...633..810M}.  NGC 4945 is one of the closest Milky Way analogs with a near solar metallically, particularly at increasing distances from the galactic center where the supernova is located \citep{2015arXiv150802754S}.  This is of importance, particularly in dust formation, as higher metallicity galaxies tend to be able to produce higher dust masses \citep{2007ApJ...663..866D, 2011A&A...532A..56G}. On 2011 December 19, optical spectra obtained of the object indicated that it was a Type II supernova which most closely matched SN 2004et about a week after maximum \citep{2011CBET.2946....1M}, although maximum optical light of SN 2011ja does not appear to have been achieved until $\sim$ 2012 January 14 (Figure 2). Radio observations presented in  \cite{2013ApJ...774...30C}, further suggest an explosion date of 2011 December 12 (JD 2455907). Therefore, throughout this paper we choose 2011 December 12 as day 0 for SN 2011ja, and day 34 as the date of maximum light.  In Section 2, we discuss the data reductions and present a comprehensive analysis of these data in Section 3. Section 4 contains a discussion of the implications of the data, and Section 5 includes analysis in regards to dust formation, CSM location, and progenitor characteristics.  Finally, in Section 6 we briefly summarize the significant results of the paper.

\section{Observations and Data Reduction}

Optical imaging and spectra were obtained with GMOS/Gemini South (GS-2012A-Q-79, GS-2013-Q-93, PI Andrews). A Color composite Gemini/GMOS image of SN 2011ja is shown in Figure 1, and a summary of observations and resultant photometry in Table 1. The g$^{\prime}$r$^{\prime}$i$^{\prime}$ images were reduced and stacked using the IRAF\footnote[1]{IRAF is distributed by the National Optical Astronomical Observatory, which is operated by the Association of Universities for Research in Astronomy, Inc., under cooperative agreement with the National Science Foundation.} \textit{gemini} package.  The instrumental g$^{\prime}$r$^{\prime}$i$^{\prime}$  magnitudes were transformed to standard Johnson-Cousins VRI using tertiary standards created from stars in the field (Figure 1 and Table 2) and transformations presented in \citet{2007ApJ...669..525W}. Uncertainties were calculated by adding in quadrature the transformation uncertainty quoted in \citet{2007ApJ...669..525W}, photon statistics, and the zero point deviation of the standard stars for each epoch. Figures 2 and 3 show the resultant light curves.

For each GMOS/Gemini South epoch, three spectra of 900s were obtained in semester 2012A and six spectra of 900s were obtained in semester 2013A. The spectra were obtained in longslit mode using grating B600 and a slit width of 0$\farcs$75.  With a resolving power R = 1688, this corresponds to a velocity resolution of $\sim$ 180 km s$^{-1}$. Central wavelengths of 5950, 5970, and 5990 \AA\ were chosen to prevent important spectral features from falling on chip gaps. A 2x2 binning in the low gain setting was used. Spectra were reduced using the IRAF \textit{gemini} package.  The sky subtraction regions were determined by visual inspection to prevent contamination from material not associated with the SN, and the spectra were extracted using 15 rows centered on the SN. The spectra from each individual night were averaged and have been corrected for the radial velocity of  NGC 4945 (460 km s$^{-1}$) obtained from narrow line emission \citep{2011CBET.2946....1M}. Spectrophotometric standards were not taken as part of our Gemini program but are taken at least once per semester.  For each of our observations we have flux calibrated our spectra using the spectrophotometric standard that is closest in date to the observation. The final spectra are presented in Figures 4 and 5.

\begin{table}
  \centering
    \caption{Gemini/GMOS Photometry of SN 2011ja} 
  \begin{tabular}{ccccc} 
  \hline
Day&JD&V & R & I\\
\hline
84 & 2455991 & 17.20 $\pm$ 0.08 & 14.87 $\pm$ 0.06 & 13.76 $\pm$ 0.06 \\ 
112 & 2456019 &18.00 $\pm$ 0.10 & 15.71 $\pm$ 0.06 & 14.62 $\pm$ 0.06\\
159 & 2456066 &18.37 $\pm$ 0.05 & 16.10 $\pm$ 0.05 & 15.03 $\pm$ 0.04\\
450 &2456357 & 19.88 $\pm$ 0.10 & 18.42 $\pm$ 0.11 & 17.64 $\pm$ 0.10\\
508 & 2456415 &19.95 $\pm$ 0.10 & 18.76 $\pm$ 0.12 & 17.88 $\pm$ 0.10\\
807 & 2456714 &19.74 $\pm$ 0.11 & 18.61 $\pm$ 0.11 & 18.05 $\pm$ 0.08\\
  \hline
  \hline
  \end{tabular}
  \label{tab:booktabs}
\end{table}
\begin{table}
\caption{Tertiary VRI Standards for NGC 4945}
\centering
\begin{tabular}{cccccc}
\hline
\hline
Star&$\alpha$ (J2000) &$\delta$ (J2000)&V&R&I\\
& 196$^{h}$$+$&-49$^{\circ}$$+$&&&\\
\hline
A & .341633 & .495212 & 16.30 & 15.42 & 14.98 \\
B & .350917 & .506569 & 15.39 & 14.90 & 14.73 \\
C & .353678 & .507214 & 15.70 & 14.43 & 14.93 \\
D & .345111 & .517944 & 16.31 & 16.26 & 15.67 \\ 
E & .240586 & .501742 & 16.44 & 15.82 & 15.57 \\
F & .243756 & .514639 & 16.60 & 16.11 & 15.56 \\
G & .297883 & .493965 & 15.85 & 14.56 & 15.12 \\
H & .279450 & .516431 & 16.28 & 14.88 & 15.45 \\
I & .297790& .494104 & 16.77 & 15.81 & 16.02 \\
\hline
\end{tabular}
Photometry is from NOMAD and DENIS catalogs.
\centering
 \label{tab:vristandards}
\end{table}

Seven epochs of  $JHK_{s}$ imaging, spanning 8 to 272 days post-explosion, were obtained from NTT/SOFI data made available on the ESO archive (184.D-1140(N), PI Benetti). A summary of the epochs and resultant photometry is presented in Table 3 and shown in Figure 2. Observations were taken with a pixel scale of 0.288$^{\prime\prime}$/pixel in a four-point dither pattern for background subtraction. Images were corrected for crosstalk and background contamination using IRAF scripts provided on the SOFI website then reduced, aligned, and stacked using standard IRAF techniques. Differential aperture photometry was performed using the 2MASS standards listed in Table \ref{tab:2mass}. On 2011 December 17 (Day 15) and 2012 April 30 (Day 113) we also obtained  $JHK_{s}$ imaging from the 0.8m Infrared Imaging System (IRIS) telescope at Cerro Armazones Observatory, Chile (operated by the Astronomical Institute of the Ruhr-Universitat Bochum and the Universidad Catolica del Norte.)  Aperture photometry was also performed using the standards presented in Table 4. Additional low resolution optical spectra for days 64 and 237, observed with NTT/EFOSC, were also obtained from the ESO archive (184.D-1151(Z), PI Benetti).  Observations were taken using a slit width of 1$\farcs$0 and grism 13, which has a resolution of $\sim$ 2.8 \AA/pixel.  Standard IRAF reduction procedures were applied, and background subtraction was done using the same area as for the GMOS spectra.  Flux calibration was achieved using spectrophotometric standards taken on the night of observation.

Spitzer IRAC (3.6 and 4.5 $\mu$m) images were obtained at four epochs from day 105-857. The images were mosaicked and resampled using standard MOPEX procedures to improve photometric quality. A summary of the observations and fluxes are shown in Table 3.  PSF photometry was performed using the Point Response Functions (PRF) developed for IRAC. Additional mid-IR observations at 18.72 and 10.77 $\mu$m  were obtained with VLT/VISIR (288.D-5031(A), PI Wesson) on 2012 February 26 (day 76)  and March 14 (day 93), respectively .  For the B10.7 filter a total 3600s was taken on target as was 7200s for the Q2 filter. Both observations resulted in a non-detection, so only upper limits can be inferred. These are 2 mJy for 10.7 $\mu$m and 10 mJy for 18.72 $\mu$m.
\begin{table*}
  \centering 
   \caption{Near- and Mid- IR Photometry of SN 2011ja}
  \begin{tabular}{ccccccc} 
  \hline
Day&JD&J & H & K& 3.6 $\mu$m (mJy)& 4.5 $\mu$m (mJy)\\
\hline
8 & 2455915 & 11.46 $\pm$ 0.05 & 11.85 $\pm$ 0.03 & 10.76 $\pm$ 0.06&& \\ 
15 & 2455922 & 11.12 $\pm$ 0.04 & 10.93 $\pm$ 0.04 & 10.32 $\pm$ 0.04&& \\
40 & 2455947 &10.78 $\pm$ 0.03 & 10.29 $\pm$ 0.03 & 9.92 $\pm$ 0.06&&\\
65 & 2455972 &11.15 $\pm$ 0.03 & 10.63 $\pm$ 0.02 & 10.20 $\pm$ 0.07&&\\
93 &2456000 & 12.59 $\pm$ 0.03 & 12.02 $\pm$ 0.04 & 11.60 $\pm$ 0.06&&\\
105 &2456012 & - & -&- & 12.54 $\pm$ 0.29&14.35 $\pm$ 0.24\\
113 &2456020 & 12.85 $\pm$ 0.10 & 12.22 $\pm$ 0.10 & 11.60 $\pm$ 0.15&&\\
121 & 2456028 &13.12 $\pm$ 0.02 & 12.45 $\pm$ 0.04 & 11.87 $\pm$ 0.06&&\\
243 & 2456150 &14.11 $\pm$ 0.11 & 12.65 $\pm$ 0.07 & 11.49 $\pm$ 0.07&&\\
272 & 2456179 &14.39 $\pm$ 0.04 & 12.91 $\pm$ 0.02 & 11.63 $\pm$ 0.04&&\\
486 &2456393 & - & -&- & 13.48 $\pm$ 0.29& 15.32 $\pm$ 0.24\\
637 &2456544 & - & -&- & 10.16 $\pm$ 0.26& 11.63 $\pm$ 0.22\\
857 &2456764 & - & -&- & 7.79 $\pm$ 0.20& 8.17 $\pm$ 0.17\\
  \hline
  \hline
  \end{tabular}
  \label{tab:booktabs}
\end{table*}

\section{Analysis}

\subsection{Internal Extinction}
The initial discovery spectrum, shown in Figure 4, suggested a large amount of extinction toward SN 2011ja \citep{2011CBET.2946....1M}. The Milky Way foreground is only A$_V$=0.48 mag \citep{2011ApJ...737..103S} so much of the extinction is 
internal to NGC 4945, which is not unexpected as extinction estimates along its galactic plane are A$_{v}$ $>$ 11-13 \citep{2000A&A...357...24M}.  In order to determine an accurate extinction value for SN 2011ja, we have employed a variety of tests.  Because of the low recession velocity of the galaxy (450 km s$^{-1}$) the use of Na ID doublet was insufficient for this galaxy due to the blending of internal and external Na absorption lines.  This is not necessarily a hindrance, as it is still under debate whether this method is acceptable for SNe \citep{2011MNRAS.415L..81P, 2012MNRAS.426.1465P}, especially in the presence of circumstellar material.  Therefore, we compared our unreddened, early-time spectra with optical spectra of SN 2004et, a prototypical Type IIP, from similar epochs and applied extinction corrections until the spectra were coincident.  SN 2004et has an E(B-V) = 0.43 \citep{2006MNRAS.372.1315S}, and comparisons on both day 7 and day 84 yield a total E(B-V)=1.8 in SN 2011ja, using the reddening law of CCM \citep{1989ApJ...345..245C}.  Throughout this paper we use a value of total reddening, foreground to SN2011ja, to be E(B-V) =1.8.

\begin{table}
\caption{Tertiary 2MASS Standards for NGC 4945}
\centering
\begin{tabular}{cccccc}
\hline
\hline
Star&$\alpha$ (J2000) &$\delta$ (J2000)&J&H&K\\
& 196$^{h}$$+$&-49$^{\circ}$$+$&&&\\
\hline
1 & .265558 & .507378 & 12.68 & 12.44 & 12.38 \\
2 & .279346 & .516434 & 14.77 & 14.35 & 14.27 \\
3 & .264956 & .535126 & 15.08 & 14.64 & 14.55 \\
4 & .316055 & .519047 & 15.79 & 15.01 & 14.87 \\
5 & .287417 & .534851 & 15.50 & 14.90 & 14.84 \\
\hline
\end{tabular}
\centering
 \label{tab:2mass}
\end{table}

  As a second method and sanity check, we measured the relative line strengths of the H$\alpha$ and H$\beta$ nebular emission, assuming both Case A and Case B recombination in the HII region surrounding the SN, as these are the two extremes of hydrogen recombination, and the area surrounding SN 2011ja must lie somewhere in between.  In the absence of reddening we would expect a ratio of 2.85 for Case B and 1.91 for Case A, but we find a ratio closer to 17.25 suggesting  1.64 $<$ E(B-V) $<$ 2.01 \citep{1979MNRAS.187P..73S}. This is in reasonable agreement with comparison to the SN 2004et spectra. When we measure this ratio on day 159, E(B-V) has increased to 2.10, indicating an additional increase in A$_v$ of 0.55 mag between these two dates, assuming R$_v$ = 3.1.  As we will discuss below, this is likely due to early-dust formation.
  \begin{figure*}
   \centering
   \includegraphics[width=5.5in]{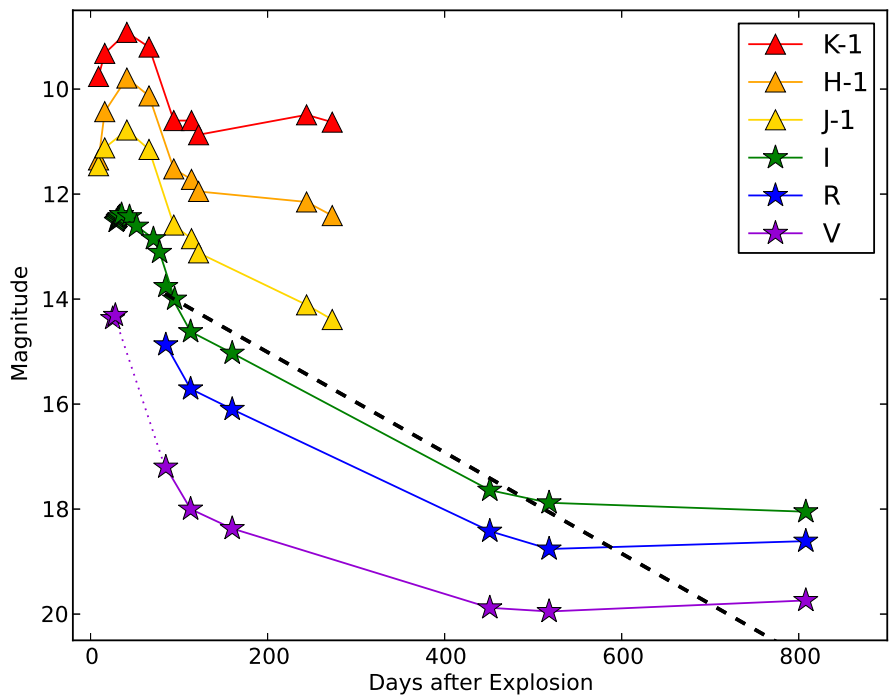} 
   \caption{Optical and IR light curves of SN 2011ja from values listed in Tables 1 and 3. The NIR curves have been shifted up one magnitude for clarity.  V and I photometry prior to day 84 has been obtained from AAVSO.  The dashed line follows the decay of $^{56}$Co.}
   \label{fig:example}
\end{figure*}
\subsection{Optical Lightcurve Evolution}

Figure 2 shows the VRI evolution of SN 2011ja between days 28 - 1138.  The I-band AAVSO observations up to day 77 indicate that the SN may have been increasing in luminosity since discovery and reached a maximum brightness on 2012 January 14, day 34.  Shown in detail in Figure 3, the plateau phase only lasted for $\sim$ 35 days, with a decline rate of 0.01 mag day$^{-1}$.  The drop into the radioactive decay portion of the light curve that begins on day $\sim$70, lasted only 15 days with a $\Delta$I of 1 magnitude, which was similar for the other optical bands.  It is possible that the transitional phase lasted until day 112, and if this is the case, the SN faded by 1.8 mags in 42 days (a rate of 0.04 mag day$^{-1}$).  We have reason to believe, as explained below, that the start of the radioactive decay phase was in fact around the time of our first GMOS observation of SN 2011ja on day 84.

Comparison with the lightcurves of other Type II SNe shown in Figure 3, particularly after day 50,  shows that SN 2011ja may lie somewhere between a Type IIP and a Type IIL classification-wise.  The blurring between ``plateau" and ``linear"  designations of Type II SNe is becoming more common as the sample of well studied objects increases, with a survey of hydrogen rich SNe presented by \citet{2015PASA...32...19A} concluding that there are not two distinct classes, but rather a continuum based most strongly on the envelope mass at explosion.  This is somewhat at odds with \citet{2012ApJ...756L..30A} who make a contrary claim that there are distinct groups based on progenitor types.  The short plateau duration, the absolute magnitude at maximum 
(I = -18.3), and the steep drop into the radioactive decay phase all point to a CCSN with a smaller hydrogen envelope, which we will discuss in more detail below.

Assuming the radioactive decay phase started on day 84, the V-band magnitudes should then be 17.47 on day 112 and 17.92 on day 159 if the only energy source is the decay of $^{56}$Co (dashed line in Figure 2). This is roughly 0.5 magnitudes brighter than the observed magnitude, requiring an increase in optical depth ($\tau$) of 0.5 between day 84 and 112 to account for the magnitude deficit.  Modeling of the optical and IR photometry lend credence to this hypothesis, and as we will show in Section 4, radiative transfer modeling shows that there is a substantial increase in $\tau_V$ between day 105 and day 486.  By itself this does not necessarily confirm an increase of extinction prior to day 159, but in tandem with the optical spectroscopy presented in the next section, we have sufficient reason to believe new dust was being formed within 100 days of explosion in SN 2011ja.

Alternatively if the radioactive decay phase is assumed to begin on day 112, the SN does not suffer from any increased extinction and behaves as expected until $\sim$400 days, when the luminosity levels out through our last observation on day 857.  This late-time plateau can be caused by various physical mechanisms, notably a scattered-light echo, shock-interaction, or ambient light from the parent stellar cluster.  These scenarios are discussed in detail in Section 5.2.

\begin{figure*} 
   \centering
   \includegraphics[width=3.4in]{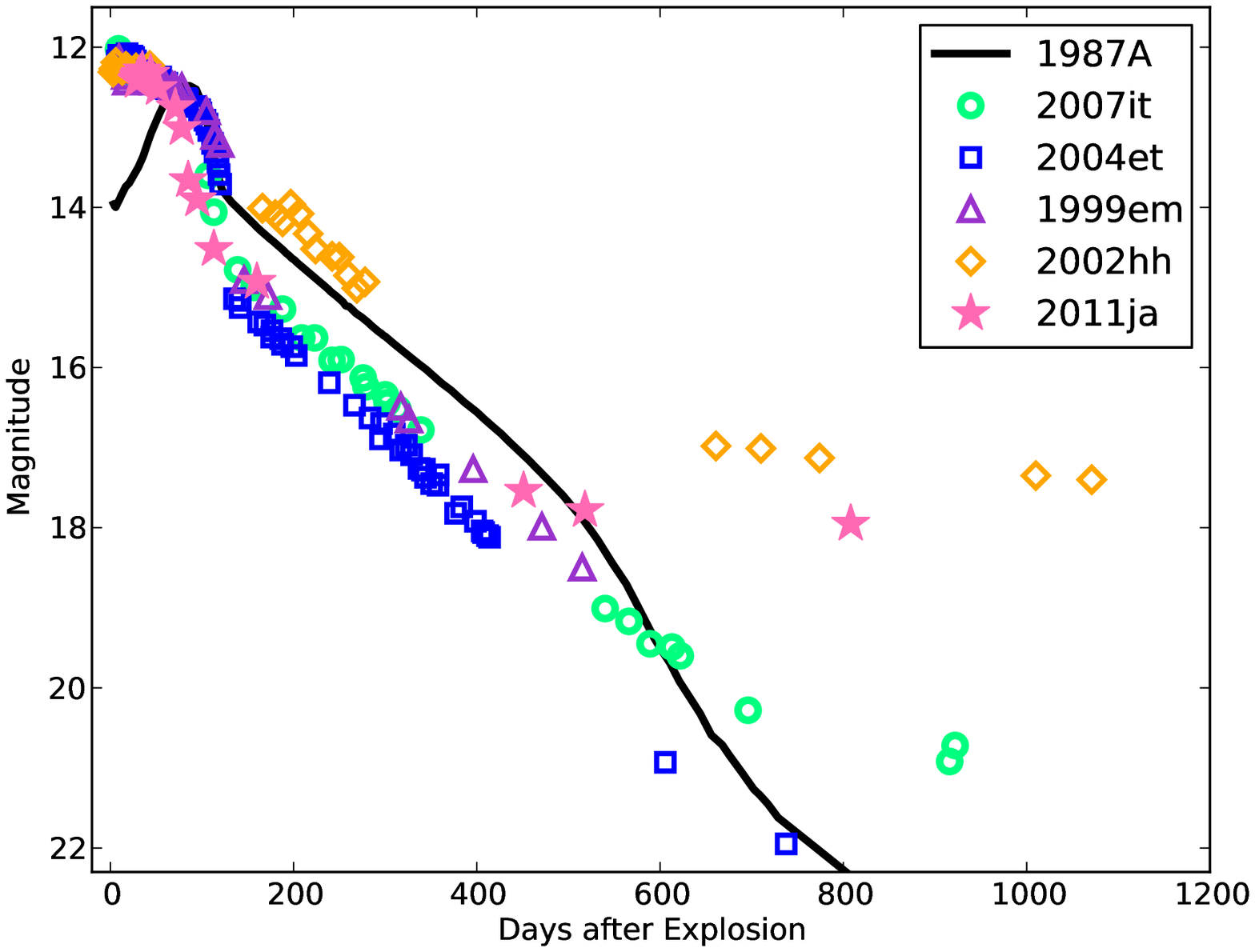} 
   \includegraphics[width=3.4in]{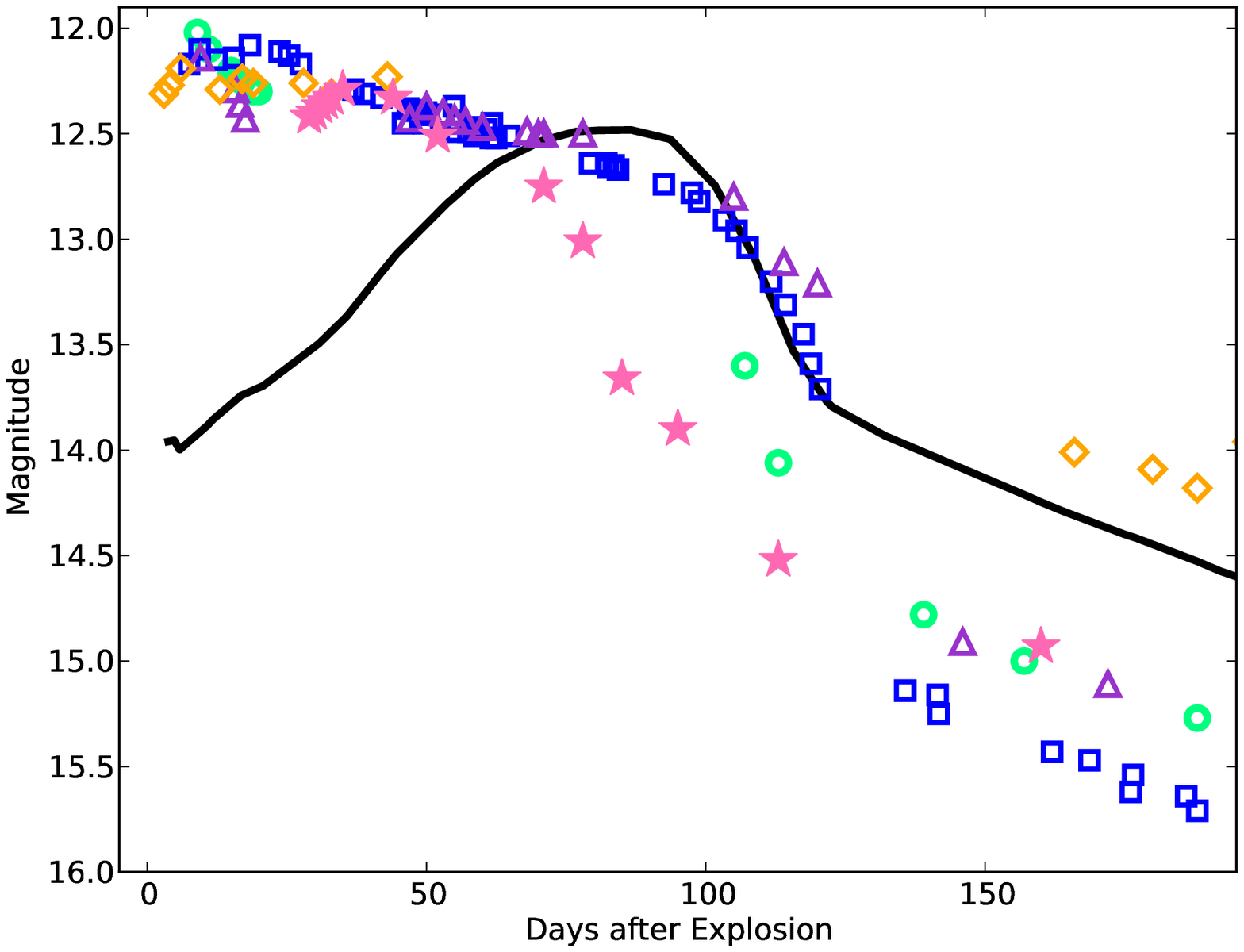} 
   \caption{Optical I-band lightcurves for a sample of Type II SNe.  The right figure shows a zoomed-in region of the first 200 days. Data are from: SN 1987A \citep{1988AJ.....95...63H,1988AJ.....96.1864S,1990AJ.....99.1146H}, SN 2007it  \citep{2011ApJ...731...47A}, SN 2004et \citep{2006MNRAS.372.1315S}, SN 1999em  \citep{2003MNRAS.338..939E}, and SN 2002hh \citep{2006MNRAS.368.1169P} }
   \label{fig:example}
\end{figure*}

\begin{figure*} 
   \centering
   \includegraphics[width=5in]{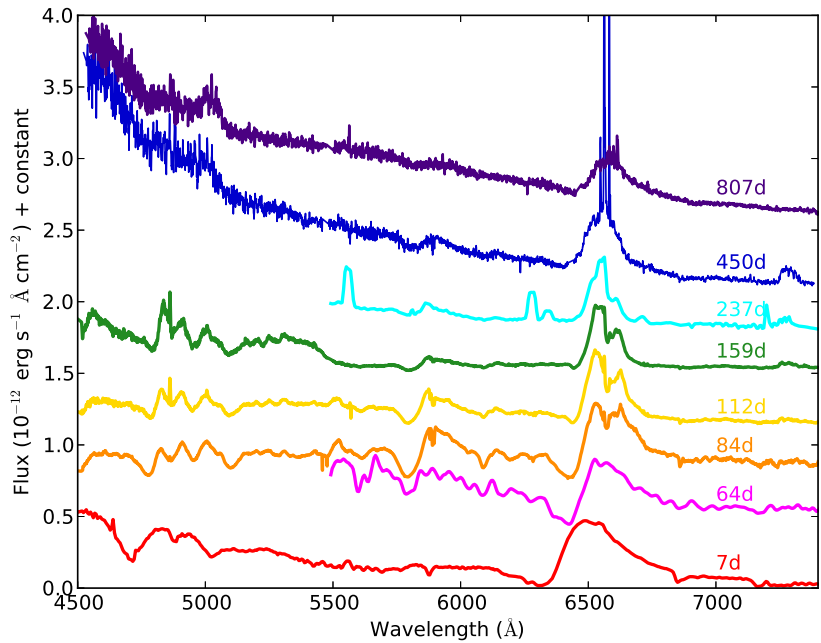} 
   \caption{Optical spectra of SN 2011ja.  Day 7, 64, and 237 are from NTT/EFOSC2, and the rest are from Gemini/GMOS.  All spectra have been flux calibrated but shifted by a constant for presentation. An extinction correction of  E(B-V) = 1.8 has been applied to all spectra. }
   \label{fig:example}
\end{figure*}

\subsection{Spectra}
Early-time spectra of SN 2011ja show a prototypical Type II supernova (Figure 4, Day 7), with blue-shifted H$\alpha$ and H$\beta$ emission lines, peaking at -3500 km s$^{-1}$ and -3000 km s$^{-1}$, respectively, and very little else.  While most other SNe show a very blue continuum at these early times, the high reddening towards SN 2011ja represses this feature. This is further exemplified by a strong Na I D absorption feature present around 5900\AA. A high-velocity feature (HV) may also be present in the Day 7 spectrum at $\sim$13,700 km s$^{-1}$, likely due to flash-heated unshocked pre-explosion mass loss from the progenitor star \citep{2007ApJ...662.1136C}. 

Our first GMOS spectra of SN 2011ja was obtained on day 84, when the multi-component hydrogen lines that seem to appear in the day 64 spectra are now prevalent (Figure 5).  Blue-shifted and red-shifted peaks are seen in H$\alpha$ at -1400 km s$^{-1}$, and 2900 km s$^{-1}$, in addition to the narrow nebular line at 0 km s$^{-1}$.  There is also a possible tertiary peak on the red side at 1700 km s$^{-1}$. The H$\beta$ emission shows peaks at -2100 km s$^{-1}$ and 3100 km s$^{-1}$, but not the existence of a second red-shifted component.   We cannot ascertain if a double-peaked structure exists in the He I $\lambda$5876 line due to the blending with the Na ID doublets from local and host galaxy extinction. Multipeaked asymmetric hydrogen lines have been seen in other SNe, such as SN 1993J \citep{2000AJ....120.1487M}, SN 1998S \citep{2000ApJ...536..239L}, SN 2004dj \citep{2007ApJ...662.1136C}, SN 2007od \citep{2010ApJ...715..541A}, SN 2009ip \citep{2014MNRAS.442.1166M}, and PTF11iqb \citep{2015MNRAS.449.1876S}, and are mostly attributed to a torrodal or disc geometry of surrounding CSM material. Comparison with models presented in \cite{2005ApJ...622..991F} indicate that the CSM around SN 2011ja may be represented by a torus inclined between 30-60 degrees with an angular thickness between 30 and 50 degrees.

\begin{figure} 
   \centering
   \includegraphics[width=3.5in]{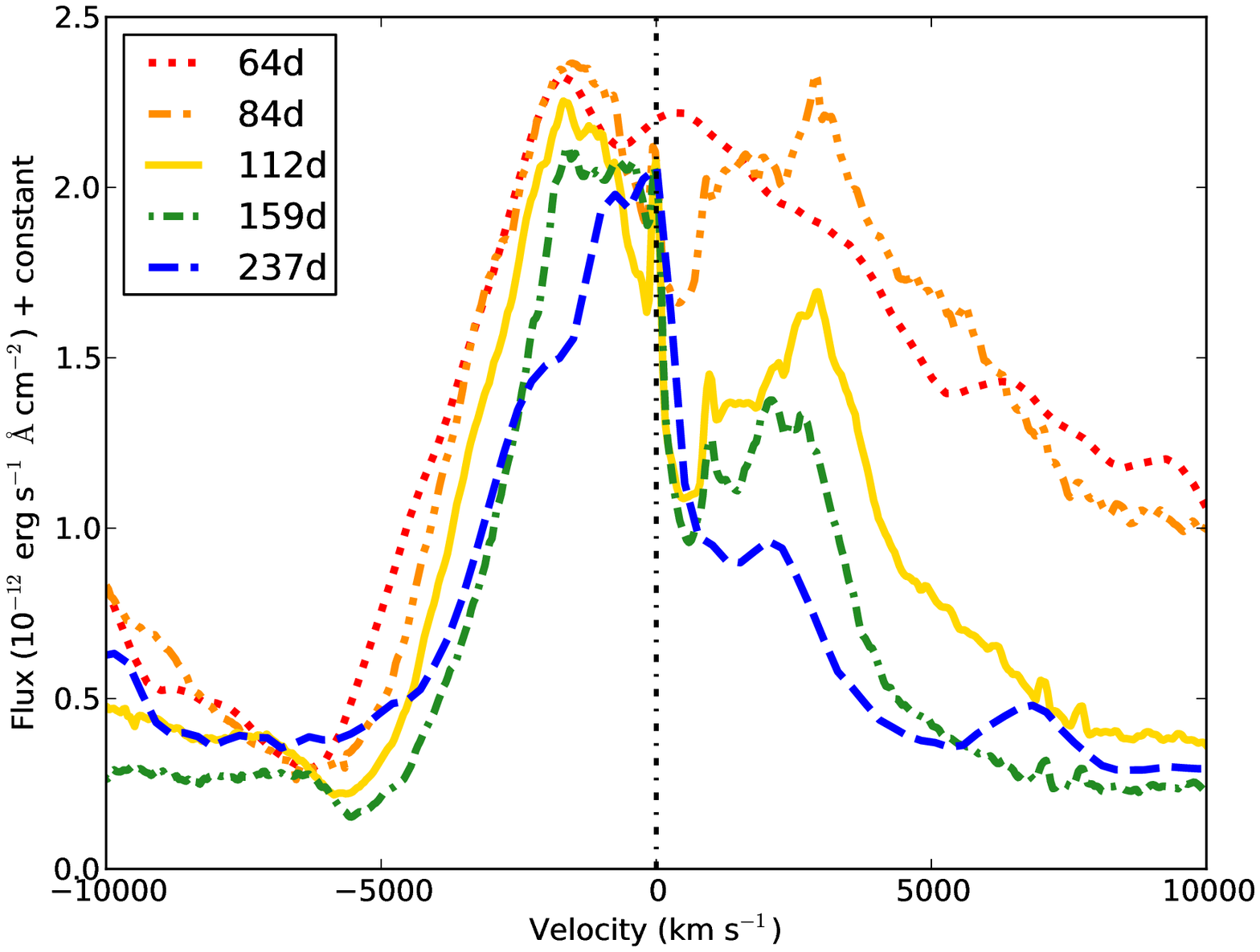} 
    \includegraphics[width=3.5in]{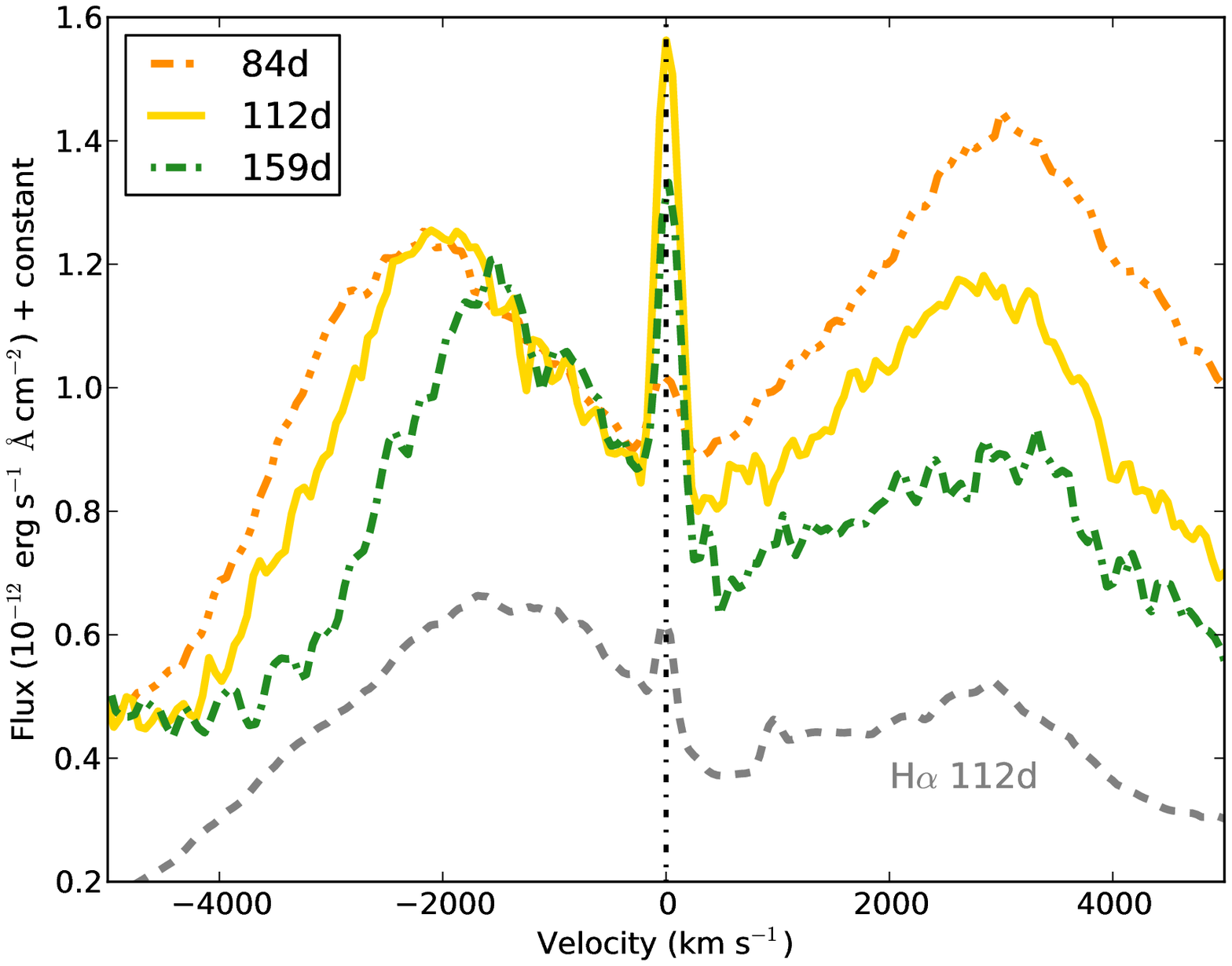} 
   \caption{ H$\alpha$ (top) and H$\beta$ (bottom) evolution for a subset of epochs from Figure 4.  The enhanced red emission in H$\beta$  on day 84 could be due a blend with Fe $\lambda$ 4924 \AA. }
   \label{fig:example}
\end{figure}

 Other broad nebular lines such as Ba II $\lambda$6142, [Sc II] $\lambda$5527, $\lambda$5658, and $\lambda$6246, and [O I] $\lambda$$\lambda$6300,6363 are dominant in the nebular spectra.  Fe II  $\lambda$$\lambda$ 4924,5169 may also be present and somewhat blended with H$\beta$ and other Sc lines.    An intermediate-width component of [O III] $\lambda$ 5007 \AA\  seems to appear at this time as well, and persists until our last spectra on day 807. Alternatively, this could be a  -735 km s$^{-1}$  blue-shifted peak of Fe $\lambda$ 5018 \AA. If this is the case, then the red emission peak of H$\beta$ is likely blended with a similarly blue-shifted component of Fe $\lambda$4924 \AA. The only other published example of broad 
 [O III] emission is in the well-studied Type IIn 1995N \citep{2002ApJ...572..350F}.  In that object, lines of a similar width (1500 km s$^{-1}$) were seen, the only difference being in SN 1995N they showed a much more boxy profile, similar to what we see in our H$\alpha$ emission lines.  Those authors attributed this emission to unshocked ejecta composed of oxygen core material coupled with a low hydrogen envelope mass.

By days 159 and 237, when the SN is well into the nebular phase, the strong P-Cygni profile in H$\alpha$ has all but disappeared. In addition to having multiple peaks, the H$\alpha$ (and to some extent H$\beta$) lines begin to show signs of flattening as the nebular phase persists.  Boxy, flat-topped spectra have also been seen in SNe 1993J \citep{2000AJ....120.1487M}, 1998S \citep{2000ApJ...536..239L}, 2004et \citep{2009ApJ...704..306K}, and 2007od \citep{2010ApJ...715..541A}, and are attributable to CSM interaction. Also noticeable in the H$\alpha$ emission is the degradation of the red peak from the day 64 and 84 observations (Figure 5).  Over the course of $\sim$180 days, the ratio of the blue and red peak emission has evolved from 1:1 to 2:1. Further comparison between H$\alpha$ emission of SN 2011ja at $\sim$day 84, SN 2004dj at day 128, and SN 2007od at day 232 (Figure 6) shows similar structures, particularly in the peaks at $\pm$1500 km s$^{-1}$.  This double-peaked structure was attributed to the interaction of the ejecta with a torus or disc of material surrounding the supernova, with the stronger attenuation on the red side due to dust forming in the line of sight. The extra peak at +3200 km s$^{-1}$ is more of a mystery. It is possible that the multiple peaks seen in the hydrogen lines could come from asymmetries in $^{56}$Ni in the ejecta.  This non-uniformity could cause  uneven ionization and excitation in the ejecta, as was suggested for the He I lines in the Type IIb SN 2008ax \citep{2011MNRAS.413.2140T}.

\begin{figure} 
   \centering
    \includegraphics[width=3.5in]{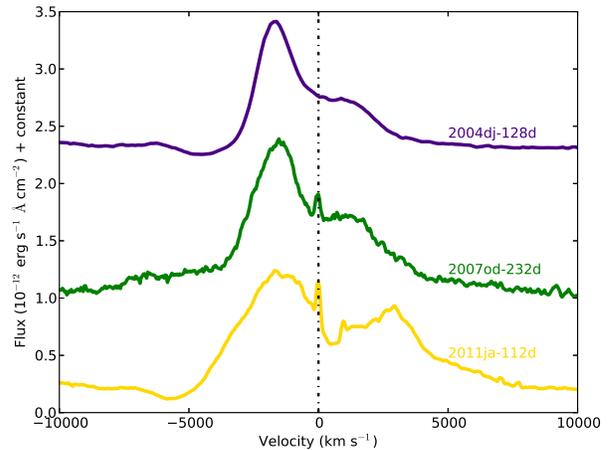}
   \caption{ H$\alpha$ profiles of other SNe with known CSM interaction.  Data are from \citet{2010ApJ...715..541A} (2007od) and \citet{2006MNRAS.369.1780V} (2004dj).  All three SNe were classified as normal Type II, yet all three show the same double-peaked emission profile, with a dominant blue-peak within 4-8 months post-explosion.}
   \label{fig:example}
\end{figure}

The H$\alpha$ line width of 5000 km s$^{-1}$ persists from day 64 until our last observation on day 807.  Late-time spectra show the emergence of strong lines of [N II] $\lambda\lambda$ 6548,6583 and [S II] $\lambda\lambda$ 6716,6731 \AA  (see Figure 4).  More striking is the blue continuum that emerged in the last two epochs. This occured at the same time as the leveling off of the optical lightcurve discussed in section 3.2.  The two are likely caused by the same phenomenon, and are discussed in detail in Section 5.3.

\subsection{IR Observations}
In Figure 2 we also show the first $\sim$ 300 days of JHK evolution, including a maximum around day 40.  Between day 8 and 40 there is an increase in the absolute magnitude of all NIR bands, although the largest increase of 1.5 mag occurred in the H-band.  In the weeks following the maximum there is a steep decline, very similar to the optical lightcurve evolution, until day 121.  Between day 121 and 243 the K-band brightens by $\sim$ 0.4 magnitudes and an almost constant plateau in the H-band.  This corresponds to the same time period when we believe there was an increase in optical fading by 0.5 magnitudes, and that the H$\alpha$ emission showed the most red-side attenuation.   A similar increase in the H- and K-bands was observed around day 50 in SN 2005ip \citep{2009ApJ...691..650F} and SN 2006jc \citep{2008ApJ...680..568S}. In both cases, the increase in the NIR can be attributed to dust grain formation.  Although there is no NIR photometry available to us after day 272, the brightness is once again declining into a radioactive tail, indicating that the new grains have either been destroyed in the reverse shock or have cooled sufficiently to only be detectable at longer wavelengths.

\subsection{$^{56}Ni$ Mass Estimates}
We estimate a $^{56}$Ni mass for SN 2011ja of  M$_{Ni}$= 0.22$\pm$0.03 M$_{\sun}$ employing the methods of \citet{2003ApJ...582..905H}, using the V magnitude at day 84. The bolometric luminosity of the radioactive tail (in erg s$^{-1}$) is calculated as 
\begin{equation}
log_{10}L_t= \frac{-[V_t-A(V)+BC] + 5log_{10}D-8.14}{2.5},
\end{equation}
with A$_{V}$ = 5.6 mag, and a bolometric correction of BC = 0.26. The nickel mass is then calculated at various times during this phase, as
\begin{equation} 
M_{Ni} = (7.866 x 10^{-44})L_texp[\frac{\frac{t_t-t_0}{1+z}-6.1}{111.26}]M_{\sun}.
\end{equation}

Here 6.1 days is the half-life of $^{56}$Ni and  11.26 days is the e-folding time of $^{56}$Co and $t_{t}$-$t_{0}$ is the age of the SN. This calculated Ni mass is much larger than other Type IIP SNe.  There is, of course, some uncertainty in the extinction correction, not to mention the luminosity at each epoch could be underestimated due to additional dust formation or overestimated due to circumstellar interaction.  If this is an accurate measurement of the $^{56}$Ni mass, it would put SN 2011ja in a regime of high mass normally reserved for stripped envelope CCSNe, particularly IIb and Ib \citep{2014arXiv1406.3667L}. For comparison, other ``normal" Type IIp SNe such as  SNe 1999em, 2003gd, and 2004dj each have $^{56}$Ni masses $\sim$0.02 M$_{\sun}$, or a full order of magnitude lower than estimated here \citep{2003MNRAS.338..939E,2005MNRAS.359..906H,2006MNRAS.369.1780V}.

\section{Radiative Transfer Modeling}
To determine the amount and type of dust being formed in SN 2011ja, we have used the MOCCASIN 3D Monte Carlo radiative transfer code \citep[and references therein]{2005MNRAS.362.1038E}. As was done for SN2007it \citep{2011ApJ...731...47A} and SN 2010jl \citep{2011AJ....142...45A} we assume three different dust geometries.  The ``smooth'' distribution describes dust distributed uniformly  within a spherical shell surrounding the SN.  The ``torus'' model distributes dust uniformly within a torus at some inclination around the SN.  Finally, the ``clumpy'' model places dust in clumps and scatters it throughout a spherical shell surrounding the SN.  For grain sizes we are using a standard MRN grain size distribution of $a^{-3.5}$ between 0.005 and 0.05 $\mu$m \citep{1977ApJ...217..425M}.  For each epoch, we use the Spitzer IRAC data to set the date, and extrapolate the optical and NIR photometry from the surrounding observations.  We also correct all photometric points for the assumed E(B-V)=1.8.  

Continuing on in the manner presented in Andrews et al. (2010, 2011a,b), the dust and luminosity for the source was located between an inner radius R$_{in}$ and an outer radius R$_{out}$ of a spherically expanding shell, with the diffuse emission luminosity being proportional to the density at each location.  The smooth model assumes the density of dust in the shell was inversely proportional to the square of the radius. For the clumpy model the photons originate in the inhomogeneous interclump medium, where the clumps are considered to be optically thick and spherical.  For the torus models, densities are specified for the inner and outer walls, with the dust distribution falling off linearly between the two radii \citep{2007MNRAS.375..753E}. For each model we used grains of amorphous carbon (AC) using the optical constants of \citet{1988ioch.rept...22H}, due to the lack of mid-IR detection at 10.8 $\mu$m, which would have indicated a strong silicate component.  Inputs for the models were luminosity, ejecta temperature, inner and outer radii, and dust masses. For each epoch and parameter, initial estimates were accomplished using blackbody fits to the optical and IR data (solid curves Figure 7). This yielded dust temperatures (T$_{d}$) of roughly 550 K for the first epoch (day 105), and 725K for the remaining epochs. These temperatures are consistent with typical warm dust temperatures. Inputs and outputs for each epoch and geometry are listed in Tables 5 and 6.
\begin{figure*}
   \centering
    \includegraphics[width=3.4in]{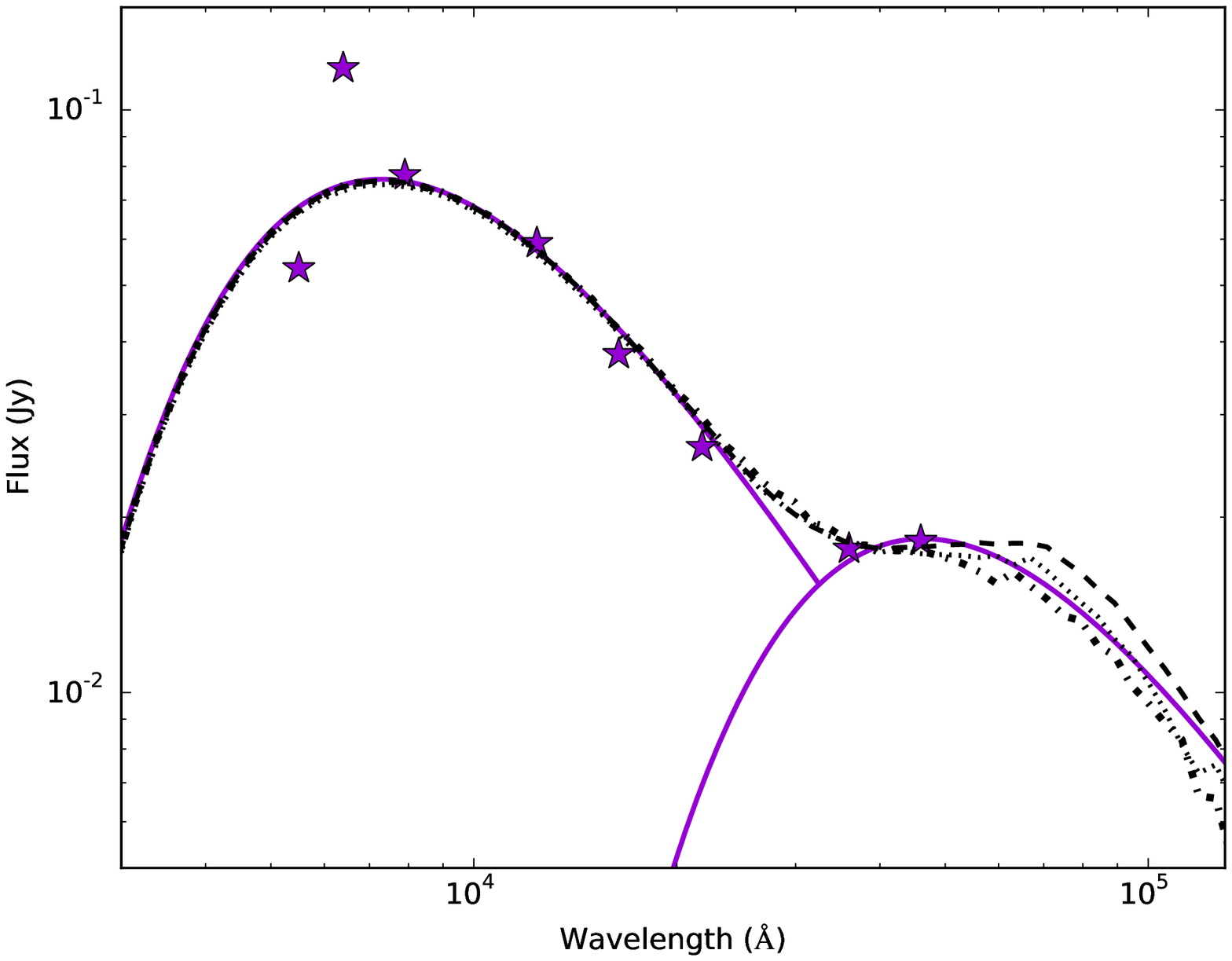}
    \includegraphics[width=3.4in]{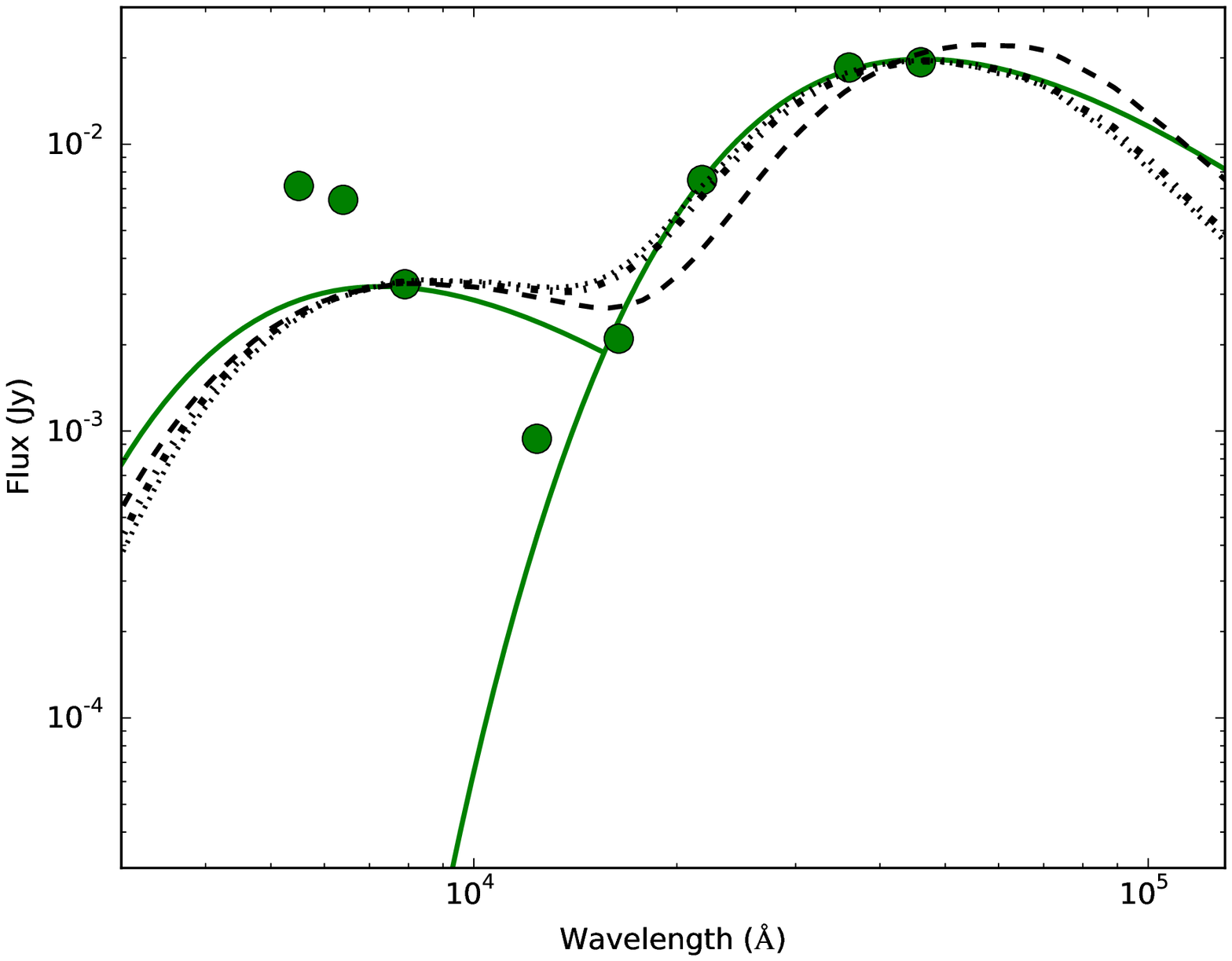}
    \includegraphics[width=3.4in]{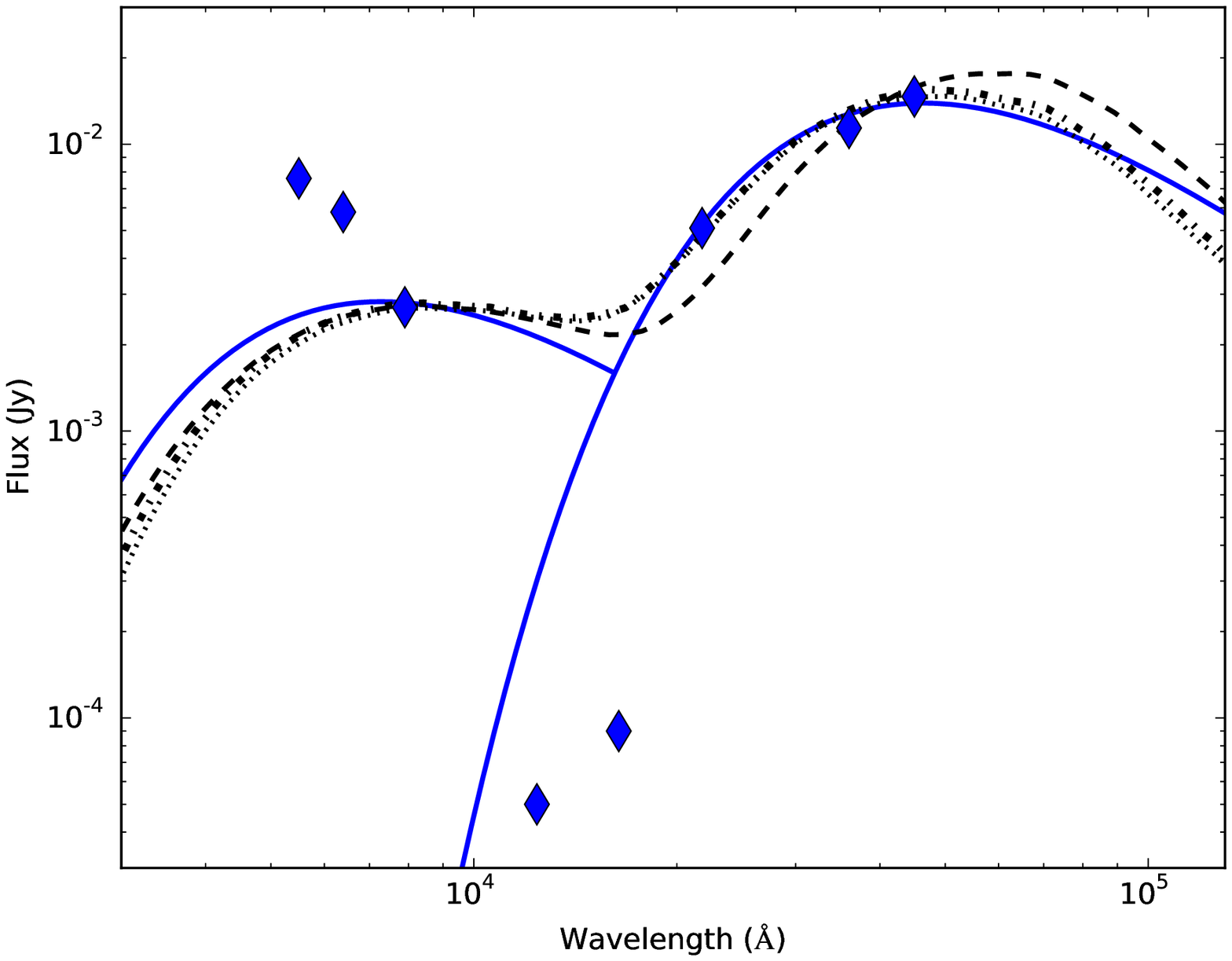}
    \includegraphics[width=3.4in]{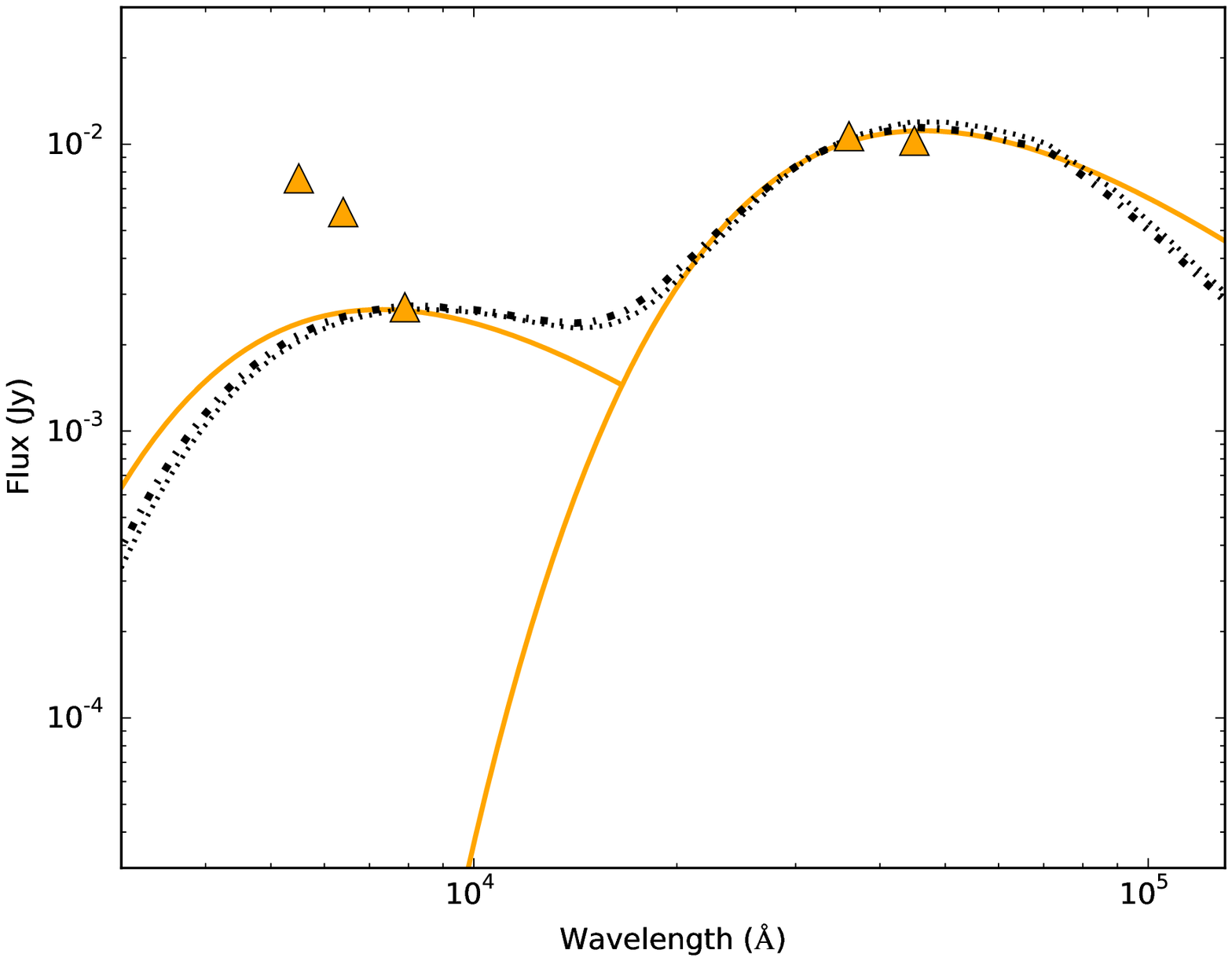}
   \caption{ SEDs of days 105 (upper left), 486 (upper right), 637 (lower left), and 857 (lower right).  All photometric points have been corrected for E(B-V)=1.8, and optical and NIR fluxes have been extrapolated from other observations to match the Spitzer observation dates. Blackbody fits are shown as solid lines. Mocassin fits are shown in dashed (smooth), dotted (torus, 45$^{\circ}$), and dash-dotted (clumpy).  Note the y-scale is different for Epoch 1 than for the other Epochs.}
   \label{fig:example}
\end{figure*}

\begin{table*}
\centering
\caption {Monte Carlo Radiative Transfer Shell Models}
\begin{tabular}{lcccccccc}
\hline
\multicolumn{2}{c}{} &
\multicolumn{5}{r}{Smooth} &
\multicolumn{1}{c}{} &
\multicolumn{1}{c}{Clumpy} \\
Epoch & T$_{ej}$ (K) & R$_{in}$ (cm) & R$_{out}$(cm) & L$_{tot.}$ (L$_{\odot}$) & $\tau$$_{v}$ & M$_{d}$ (M$_{\odot}$) & & M$_{d}$ (M$_{\odot}$) \\
\hline
105 d & 7000 &5.3e16 &5.3e17  & 1.8e8 & 0.03 & 1.0e-4 & & 1.9e-4 \\
486 d &  7000 &4.5e15 & 4.5e16 & 1.3e7 & 2.60 & 6.8e-5 & & 1.4e-4 \\
637 d & 7000 & 4.5e15 & 4.5e16 & 1.0e7 & 2.34 & 6.1e-5 & &1.2e-4  \\
857 d &  7000 & 3.8e15 &4.2e16 & 9.1e6 & 1.76 & 3.9e-5 && 8.6e-5   \\
\hline
\hline
\end{tabular}
\end{table*}

\begin{table*}
\centering
\caption {Monte Carlo Radiative Transfer Torus Models}
\begin{tabular}{lccccccccccc}
\hline
\multicolumn{5}{c}{} &
\multicolumn{2}{c}{Face-On (0$^{\circ}$)} &
\multicolumn{2}{c}{45$^{\circ}$} &
\multicolumn{2}{c}{Edge-On (90$^{\circ}$)} &
\multicolumn{1}{c}{}  \\
Epoch & T$_{ej}$ (K) & R$_{in}$ (cm) & R$_{out}$(cm) & L$_{tot.}$ (L$_{\odot}$) & $\tau$$_{v}$ & M$_{d}$ (M$_{\odot}$) & $\tau$$_{v}$ & M$_{d}$ (M$_{\odot}$) &$\tau$$_{v}$ & M$_{d}$ (M$_{\odot}$) &  \\
\hline
105 d & 7000 &5.0e16 &5.0e17  & 1.8e8 & 0.0 & 1.3e-4& 0.03  &1.5e-4 & 0.05  & 1.8e-4 & \\
486 d &  7000 &9.0e15 & 4.0e16 & 1.3e7 & 0.0 &4.8e-5& 0.0 & 6.1e-5  & 3.31 &7.7e-5 &  \\
637 d & 7000 & 9.0e15 & 4.0e16 & 9.8e6 & 0.0 & 4.3e-5 & 0.0 & 5.4e-5 & 2.94 & 6.8e-5 &    \\
857 d &  7000 & 9.0e15 &4.0e16 & 8.9e6 & 0.0 & 3.5e-5 & 0.0 & 4.3e-5 & 2.37 & 5.5e-5 &    \\
\hline
\hline
\end{tabular}
\end{table*}

For the shell models, the first epoch (day 105) is best fit by an R$_{in}$ of 5.3 $\times$ 10$^{16}$ cm, which roughly corresponds to the evaporation radius of the initial flash of the SN.   Epochs 2 and 3 (day 486 and 637 respectively) used an R$_{in}$ of 4.5 $\times$ 10$^{15}$ cm, and for our final epoch (day 857) R$_{in}$ = 3.8 $\times$ 10$^{15}$ cm.  For the torus models, the best fits were achieved with slightly different values for R$_{in}$. On day 105 we assumed R$_{in}$ = 5.0 $\times$ 10$^{15}$ cm, and for the remaining epochs  R$_{in}$ = 9.0 $\times$ 10$^{15}$ cm. We also kept the temperature of the ejecta (T$_{ej}$) at 7000K for all epochs, since this is a reasonable ejecta temperature \citep{2006Sci...313..196S} and does an adequate job fitting the first epoch of optical points.  The combination of strong H$\alpha$ and [O I] emission lines in the R band relative to the continuum, and at later times the  additional blue flux  has made it unlikely that the visible photometry is consistent with any reasonable blackbody temperature.  Keeping those two values constant, we then varied the luminosity, dust masses, and for the torus models  the number density along the inner edge  to get the most accurate fits. For all epochs we found that the clumpy distribution predicted on average about a factor of two higher dust masses than the smooth and torus models.  We have also found that varying the grain sizes and exponent on the grain distribution alters the dust mass outputs.  In general increasing the minimum grain size does not alter the dust mass, increasing the exponent integer on the grain size distribution will decrease the dust mass, and increasing the maximum grain size from 0.05  $\mu$m to 1 $\mu$m will increase dust mass by roughly a factor of two.  These values are within the range of masses  created by varying the dust geometries (Table 6), which is our largest source of uncertainty.  A recent in-depth study presented in \citet{2015arXiv150900858B} finds that a MRN distribution could not reproduce the line profiles of SN 1987A at early epochs, but as the grain size distribution is not our largest source of uncertainty we retain our standard MRN distribution for consistency.

We estimate a dust mass for our first epoch of 1.0 x 10$^{-4} M_{\sun}$ for the smooth model, 1.9 x 10$^{-4} M_{\sun}$ for clumpy, and 1.4 x 10$^{-4} M_{\sun}$ for the torus.  The remaining epochs yielded smooth dust masses of 6.7 x 10$^{-5} M_{\sun}$ for day 486, 6.0 x 10$^{-5} M_{\sun}$ for day 637, and 3.8 x 10$^{-5} M_{\sun}$ for day 857. The larger dust mass from the first epoch is likely due to the flash heating of pre-exisiting CSM dust that is just outside of the shock radius, and is roughly 200K cooler than the newly formed dust in following epochs. On day 105, the maximum distance the shock could have traveled assuming a velocity of 11000 km s$^{-1}$ is  1 $\times$ 10$^{16}$ cm making this scenario entirely plausible.  Additionally R$_{in}$ for this first epoch is larger than R$_{out}$ for the remaining epochs, indicating a different location of the modeled dust.  This dust will cool and the IR echo will fade over the next year, so that by the time of Epoch 2, the SED should be dominated by newly formed ejecta dust.

Around the time of Epoch 2, the ejecta should have traveled an average distance of 2.5 $\times$ 10$^{16}$ cm (assuming a late-time expansion of 6000 km  s$^{-1}$ measured from H$\alpha$ widths.)  This places the IR emission, and therefore the newly formed dust, mostly in the CDS interior to the point of CSM interaction.  Of course this is just the SN envelope that is traveling at this velocity.  The metal-rich core material itself is likely moving much slower, closer to the 1500 km s$^{-1}$ seen in SN 1987A \citep{1993ARA&A..31..175M}, and will only be at a distance of 6.3 $\times$ 10$^{15}$ cm. The steady decline of estimated dust masses over all of the epochs is most likely due to the dust in the distant CSM responsible for the IR echo cooling, leaving only the newly formed dust as the main near-IR contributor.  It is also possible that as the reverse shock propagates in Epochs 3 and 4 it destroys the dust that was formed in the CDS between Epochs 1 and 2, or that some of the newly formed dust has cooled as well.  This is fully consistent with the picture presented in the spectral evolution.

On day 105, the MOCASSIN models predict a $\tau_{v}$ = 0.03.  Using methods presented in \citet{2009ApJ...691..650F}, we can also independently estimate the optical depth.  If we assume L$_{max}$ = 1.25 $\times$ 10$^{9}$ L$_{\sun}$ for SN 2011ja based on the lightcurve maximum,  then E $\sim$ 1.25 $\times$ 10$^{49}$ erg using the estimates of SN 2005ip and our previous estimation of SN 2007it. We can constrain the IR energy using the black body fits, which results in an L$_{bb}$  of 5.6 x 10$^{6}$ L$_{\sun}$ and therefore an IR energy over the first 105 days of 1.97 $\times$ 10$^{47}$ erg.  Inserting these numbers into  $\tau \sim \frac{E_{IR}}{E_{IR}+E}$, we calculate a $\tau$ = 0.015, consistent with our MOCASSIN fits. 

An increase in $\tau _{v}$ between epochs 1 and 2, as mentioned above, occurs coincidentally with the emission line and optical lightcurve extinction, and therefore the formation of new dust.  While the total dust mass is actually smaller in the second epoch, the volume into which it is contained is much smaller as well, since R$_{out}$ has decreased by over a magnitude between the two, requiring similar amounts of dust in a much smaller volume. This region is also almost entirely interior to the original R$_{evap}$, again pointing to newly formed grains.  In Section 3.2 we estimate an increase in A$_{v}$ of 0.5 between day 84 and 112, model results would therefore indicate that the most likely dust geometry would be in a torus inclined at roughly 45$^{\circ}$. As can be seen in Table 6, between face-on and 45$^{\circ}$, $\tau$ = 0, but increases from 0.5 to 3.31 for inclinations between 47-90$^{\circ}$. Overall, the total mass of new, warm dust formed in SN 2011ja appears to be $\sim$ 6.0 x 10$^{-5} M_{\sun}$ by day 857, which although consistent with dust masses of other CCSNe, is still considerably smaller than the amount needed to account for the dust seen at high-z \citep{2003MNRAS.343..427M}. Although in the Introduction, we have described the much larger dust masses seen at longer wavelengths for SN1987A and other SNe observed at much later epochs.

\section{Discussion}
\subsection{Circumstellar Interaction and Environment}
The expansion of the supernova shock into the surrounding environment can create X-ray emission if the CSM is dense enough.  \citet{2013ApJ...774...30C} found that the X-ray flux in SN 2011ja increased by over a factor of 4 between day 29 and day 113 which was attributed to the supernova interacting with a CSM created from a non-steady wind. As discussed in detail above, the optical spectral evolution implies that CSM interaction is occurring by day 84 and persists onward, agreeing with the X-ray observations. In particular,   \citet{2013ApJ...774...30C} hypothesize the progenitor underwent a blue supergiant (BSG) phase shortly before explosion, producing a low density cavity immediately surrounding the SN.   Other CCSNe showing an increase in X-ray emission over the first 120 days include the type IIP SN 1999em (Pooley et al. 2002) and the type Ib SN 2006jc \citep{2008ApJ...674L..85I}, and both instances were attributed to shock interaction with the CSM.

If we use the optical spectra to constrain the date of the onset of shock interaction, we can safely assume that while it was already beginning on day 64, it was fully occurring by day 84.  Assuming the initial expansion velocity of 11 000 km s$^{-1}$ was sustained, the CSM is located 8 $\times$ 10$^{15}$ cm away from the SN.  On day 113, when the X-ray observation was taken, the shock had reached a maximum distance of 10$^{16}$ cm.

It is important here to remark on the evaporation radius (R$_{evap}$), the cavity cleared by the explosion of the supernova.  Following the evaporation radii computed by \citet{2014ApJ...797..118F} for SN 2010jl, we have estimated the ranges of R$_{evap}$ from grain size and type.  Using an absolute magnitude of I = -18.3, and standard bolometric correction of 0.26, we estimate L$_{max}$ = 4.8 $\times$ 10$^{42}$ erg s$^{-1}$, or 1/6$^{th}$ the maximum luminosity of SN 2010jl.  This means that for graphite grains (which are used in our MOCASSIN models), and a T$_{eff}$ = 10$^4$, R$_{evap}$ = 2.4 $\times$ 10$^{16}$ and  9.0 $\times$ 10$^{16}$ cm for 1.0 $\mu$m and 0.001 $\mu$m, respectively.  Therefore we can reasonably assume an R$_{evap}$ = 5 $\times$ 10$^{16}$ cm.  If of course the BC or E(B-V) estimations are under- or over- estimated this radius can change.  A lower L$_{max}$ of 1 $\times$ 10$^{42}$ erg s$^{-1}$ could bring R$_{evap}$ as close as 10$^{16}$ cm, which was the R$_{evap}$ estimated for SN 2005ip \citep{2010ApJ...725.1768F} and SN 2007it \citep{2011ApJ...731...47A}.  Additionally, a clumpier CSM could also allow grains to survive closer to the SN.  Therefore, R$_{evap}$ is consistent with the CSM interaction beginning by day 84.

As mentioned above, an intermediate-width [O III] emission line of $\sim$ 3000 km s$^{-1}$ may be visible in our day 84 spectra and persist up to the last spectral observation.  According to \citet{2002ApJ...572..350F}, if the explosion is asymmetric it is possible to get [O III] $\lambda$5007 emission at early times.  The scenario described for a similar feature seen in the Type IIn SN 1995N is that the intermediate component is actually the oxygen core of the supernova.  The presence of this line also may help explain the interesting shape of the hydrogen lines.  To get high enough ionization energies an asymmetric explosion, with most of the energy occurring in the plane, is needed.  This would create a slower expansion velocity along the plane and much faster in the polar direction.  This scenario could explain the intermediate-width H$\alpha$ and H$\beta$ lines as well, which have an underlying broad component coming from the polar region.

\subsection{New Dust Formation and Location}
Over the past few years it has become apparent that there are two modes of dust formation in CCSNe. If there is strong CSM interaction occurring it is possible for dust grains to condense in the cool dense shell (CDS) between the forward and reverse shocks. This has been seen in other SNe such as SN 1998S \citep{2004MNRAS.352..457P}, SN 2004dj \citep{2011ApJ...732..109M},  SN 2005ip \citep{2009ApJ...691..650F,2009ApJ...695.1334S}, SN 2006jc \citep{2008ApJ...680..568S}, SN 2007od \citep{2010ApJ...715..541A}, and SN 2010jl \citep{2012AJ....143...17S,2014Natur.511..326G}. Taken solely on the formation of grains within the first 150 days of explosion in SN 2011ja would point to dust formation in the CDS. The appearance of asymmetrical hydrogen lines in the optical spectra is a smoking-gun.  We cannot rule out the possibility of small amounts of dust forming in the expanding ejecta at later times, but observational evidence points to new dust forming between day 84 - 112 in a region between 4.5$\times$ 10$^{15}$ and 4.5$\times$ 10$^{16}$ cm away.

\subsection{Late-time luminosity}
\begin{figure*}
   \centering
   \includegraphics[width=4.5in]{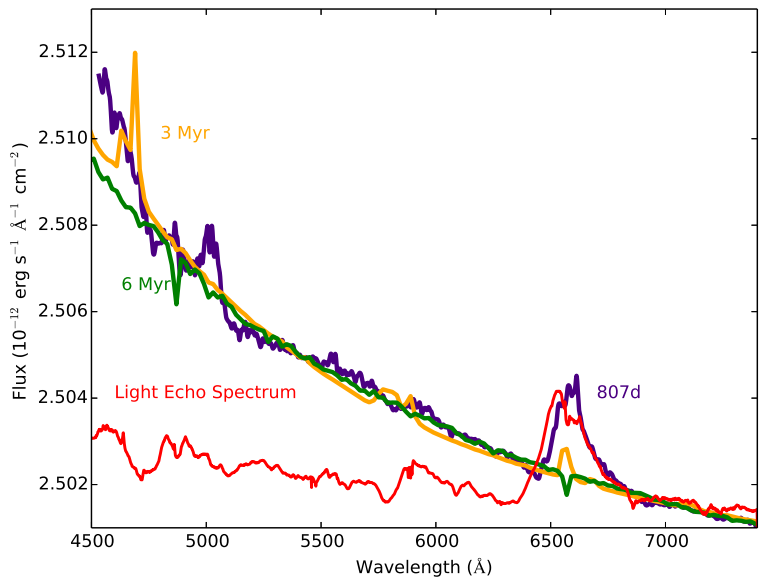} 
   \caption{ Day 807 spectrum (purple) of SN 2011ja showing enhanced blue emission.  Comparison with the light echo spectrum created from an integrated fluency of the first 84 days (red) indicates that a light echo cannot be responsible for the flux blueward of 6000\AA\ . The orange and yellow spectra are synthesized stellar populations created with Starburst99 for 3 and 6 Myr.  It is possible that the late-time luminosity has a large component of the parental stellar cluster.      }
   \label{fig:example}
\end{figure*}

After day 400 there is little change to the luminosity of SN 2011ja in VRI.  There are three possibilities to this late-time brightness: a) a scattered light echo off the surrounding CSM, b) radiative shocks, or c) fading of the SN into the ambient brightness of the parent cluster. These three possibilities are not mutually exclusive. Scattered light echoes occur when light from the supernova scatters off circumstellar or interstellar dust and redirects it back into the line of sight of the observer at some later time.  The scattering is more efficient at shorter wavelengths, causing the light echo spectrum to look bluer than the peak SN light.  At the same time, a leveling out of the optical lightcurve is created as the original light pulse travels outward through the intervening material creating a constant luminosity. The late time spectra, SED, and lightcurve all show indications that this could be a possibility, as the spectrum becomes bluer and the light curve flattens out.  The inference of pre-existing dust from the SED modeling as well as the high level of extinction in the direction of SN 2011ja also imply that there is plenty of intervening circumstellar and interstellar dust to create light echoes.  Unfortunately, when one creates a light-echo spectrum of the SN using the integrated fluency from the spectra prior to day 100, similar to that done for SN 1980K in \citet{2012ApJ...749..170S} while the emission lines appear similar, there is not enough flux in the blue to create a match (Figure 8).  So while a light echo may be present, it would be much fainter than the late-time spectrum presented here, so we have ruled this out as the cause of the late-time luminosity. A caveat to this assumption is that  the SN explosion date could be earlier than estimated here, which would mean the inclusion of an earlier, bluer light echo spectrum. Deep HST imaging of the region around SN 2011ja could resolve a scattered light echo if one exists.

The second scenario requires dense CSM to convert the kinetic energy of the expanding supernova into radiation via shocks.  This is a fairly common occurrence in Type IIn SNe at early times, and seems to be more common than originally observed in normal Type II as observations are extending past the first 3 months after explosion.  By day 807, the bulk of the SN ejecta will have reached a distance of 4.1 $\times$ 10$^{16}$ cm (or 7.7 $\times$ 10$^{16}$ for the fastest moving ejecta).  This places the edge of the ejecta clearly outside of R$_{evap}$, and at the distance of the CSM responsible for the early IR luminosity, making radiative shocks plausible. Unfortunately a shock driven luminosity, while theoretically being responsible for the constant light curve, would not make the optical spectrum bluer. Additionally, the CSM should slow the ejecta expansion, and there does not seem to be a substantial decrease in the expansion speeds in H$\alpha$ from day 450 to day 807.  Therefore we assume that radiative shocks are not responsible for the late time optical plateau, although additional X-ray observations could potentially reveal if CSM interaction is indeed occurring.

Since only $\sim$5--10$\%$ of massive stars are formed in isolation \citep{2013ApJ...768...66O}, it would not be out of the question to detect a CCSN still located in the stellar cluster from which it is formed that would become detectable as the SN faded. For instance, the type IIP SN 2004dj occurred in the massive cluster Sandage-96 and showed very similar late-time photometric and spectroscopic evolution as SN 2011ja \citep{2009ApJ...695..619V}.  The lightcurves leveled out after day 800 to the magnitude of the cluster in pre-explosion HST images.  The spectrum also showed a very blue continuum below 6000 \AA , but the nebular features of the SN (H$\alpha$, [O I]) are still present in the red part of the spectrum. Inspection of the day 807 Gemini images indicate that the size of the emitting object at the SN location is $\leq$1.$\arcsec$0, or 16 pc at the distance of SN 2011ja. Unfortunately the faintness of the object accompanied by poor seeing makes it difficult to get an accurate FWHM of the object. The average effective radii of stellar clusters are $\sim$4 pc, but cluster radii, especially for younger clusters, can extend out to 15-20 pc  \citep{2004A&A...416..537L}, making it possible that we are indeed seeing a stellar cluster+SN.  As discussed in more detail below and shown in Figure 8, comparison with Starburst99 stellar synthesis models \citep{1999ApJS..123....3L} of solar metallically  give us a decent fit with a 3-6 Myr cluster with the same intrinsic extinction of E(B-V) = 1.8.  While we cannot definitively rule out the other two options, the late-time luminosity may be largely due to the bright parental cluster of SN 2011ja.  

\subsection{Progenitor Characteristics}
Mass loss estimates for the progenitor of SN 2011ja on day 29 and 113 derived from X-ray fluxes were presented in \citet{2013ApJ...774...30C}. Using the narrow H$\alpha$ emission component present in our day 84 spectrum, we estimate the velocity of the surrounding medium to be $\sim$180 km s$^{-1}$, using the FWHM.  This is of course the limiting resolution of the spectra, suggesting that this velocity is only an upper limit.  The true value could be an order of magnitude lower and more closely resemble RSG wind speeds. This would mean a mass loss of 2 $\times$10$^{-7}$ - 10$^{-6}$ M$_{\sun}$ on day 29 and 3 $\times$ 10$^{-6}$ - 10$^{-5}$ M$_{\sun}$ on day 113, values consistent with BSG and RSG mass-loss rates respectively. For comparison HD 168625 is a BSG with measured stellar winds of $\sim$ 183 km s$^{-1}$ and $\dot{M}$ =  1.2 $\times$ 10$^{-6}$ M$_{\sun}$ \citep{1996ApJ...473..946N} and SBW1, a near twin of the progenitor of SN 1987A, has an estimated mass loss of 3 $\times$ 10$^{-7}$ M$_{\sun}$ \citep{2013MNRAS.429.1324S}. Therefore as hypothesized by  \citet{2013ApJ...774...30C}, it is likely  that the increase in X-ray luminosity could be explained by the supernova initially expanding into a low-density region then encountering a high-density RSG-like environment caused by a RSG going through a blue-loop before returning to a RSG phase. If the RSG mass-loss had a density enhancement in the equatorial region, this could also allow for easy expansion in the polar directions, while further increasing the density of the circumstellar ring via shocks.

There are other possible explanations for the presence of significant, asymmetric CSM close to the SN.  For example the two BSGs listed above, along with Sher25 are all LBV candidates with triple ring structures and more importantly, all have a central ring surrounding the star. It is possible that the environment created around these stars was caused by a violent LBV or BSG eruption, not density enhancements from RSG winds \citep{2008MNRAS.388.1127H,2007AJ....133.1034S,2013MNRAS.429.1324S}. It is also possible that the progenitor was a more massive RSG that underwent episodic mass loss, creating shells as the faster, less-dense material runs into the slower, denser material \citep{2009AJ....137.3558S, 2013MNRAS.430.1801M}. The major factor in all of these scenarios is the inferred initial mass of the progenitor. For an RSG we would expect a progenitor mass $<$ 25 M$_{\sun}$ \citep{2005ApJ...628..973L}, whereas for LBVs we could have masses as great as 40-60 M$_{\sun}$.   While originally classified as a normal Type IIP SNe, it is possible that like PTF11iqb (Smith et al. 2015), SN 2011ja was actually a Type IIn for a very brief stage.  This could allow for the possibility of a much larger progenitor, as it is generally thought that IIn SNe come from LBV or massive RSG stars. The explosion date of 2011 December 12 is only an estimate, though likely a well constrained one, and it is possible that the first observed spectrum could be multiple days older than the projected 7 days.  PTF11iqb lost the narrow IIn features $\sim$ 14 days after explosion, and any delay in its discovery would have classified it as a normal Type II.  

Figure 8 shows the day 807 spectrum fit by both a 3 Myr and 6 Myr cluster model with solar metallicity produced by Starburst99.  Without higher S/N and more accurate flux calibration it is impossible to put strict limits on the cluster age and therefore progenitor mass. Extensive stellar synthesis modeling and age-dating is beyond the scope of this paper, but the spectrum does seem to agree with a 3-6 Myr cluster and a progenitor mass $\geq$ 25 M$_{\sun}$. Using high-resolution Keck LRIS spectroscopy,  \citet{2009ApJ...695..619V} estimate that SN 2004dj had a 10-20 M$_{\sun}$ progenitor and resided in a 10-20 Myr old cluster. More recently \citet{2013AJ....146...31K} determined a cluster age of 15.6 Myr, corresponding to a mass of 14.7 M$_{\sun}$.  One notable difference is the lack of H$\alpha$ emission around Sandage-96, whereas pre-explosion H$\alpha$ imaging of NGC 4945 from the Danish 1.54m telescope indicates strong emission at the location of SN 2011ja \citep{2003A&A...406..505R}.  H$\alpha$ emission can be a strong indicator of cluster age, and is not commonly found in clusters with an age  $>$ 8 Myr \citep{2013ApJ...767...51A}.   Strong H$\alpha$ emission was also seen around SN 2010jl, and archival HST imaging may point to a natal cluster $<$ 7 Myr and progenitor mass $>$ 30 M$_{\sun}$ \citep{2011ApJ...732...63S}.  It is also interesting to note that the BSG Sher25 above may also reside in a $\sim$4 Myr old cluster \citep{2008AJ....135..878M}.

  \citet{2013ApJ...774...30C} conclude the progenitor to SN 2011ja must be a RSG with a mass $\geq$ 12 M$_{\sun}$ assuming that the X-ray variations are only due to a density enhancement by a BSG wind and that SN 2011ja was a normal Type IIP SN. The combination of parent cluster age and H$\alpha$ retention and $^{56}$Ni mass presented here, combined with previous X-ray observations all point to a larger progenitor mass for SN 2011ja than usually attributed to Type IIP SNe.  This could be of notable importance since to date, observational evidence only suggests RSG progenitor masses of Type IIP SNe of up to $\sim$16 M$_{\sun}$, the so-called ``red supergiant problem"  \citep{2009MNRAS.395.1409S,2015arXiv150402635S}.  Hypotheses for this mismatch range from underestimation of dust extinction (Walmswell $\&$ Eldridge 2012), stellar evolutionary limits (Groh et al. 2013), or direct collapse to black holes with little to no optical emission (Lovegrove $\&$ Woosley 2013, Kochanek 2014).  Smith  et al. (2011) suggest that the high mass RSGs may actually continue to evolve into blue or yellow supergiants or Wolf-Rayet stars prior to core collapse, forming other CCSNe (IIL, IIn, Ibc). This can be done either in a single or binary evolutionary track, and not only alleviates the need to invoke direct collapse to a black hole, but is also compatible with the fractional occurrence of non-IIP core-collapse events.  If in fact the progenitor of the Type IIP SN 2011ja is in the 20-30 M$_{\sun}$ range it may be one of the first definitive examples of high-mass RSGs giving rise to a CCSNe. Alternatively, the progenitor may be an LBV or a BSG and SN 2011ja may be an example of a Type II SNe discovered after a brief IIn phase.

\section{Summary}

We have presented a multi-wavelength analysis of SN 2011ja, another normal Type IIP SN, with not so normal evolution.  The short optical lightcurve plateau may place SN 2011ja into the realm of objects somewhere between IIP and IIL that have been in the spotlight in recent years. 
Signatures of dust formation are also seen after day 105, when the NIR flux increases, the H$\alpha$ emission lines become attenuated on the red side, and the optical flux decreases.  At the same time there is also a noticeable increase in the X-ray flux \citep{2013ApJ...774...30C}, which is a strong indication of CSM interaction.  This is further exemplified by the flat-topped hydrogen profiles in the optical spectra, particularly the H$\alpha$ line.  The dust formation, the flat-topped line profiles, and the increased X-ray emission all seem to occur  at roughly the same time, which paints a picture of the ejecta of SN 2011ja running into dense CSM about $\sim$4 months post-explosion.  As the ejecta runs into the CSM an area is then created between the forward and reverse shock region where $\sim$6.0 x 10$^{-5} M_{\sun}$ of dust was quickly formed.  It seems SN 2011ja is likely another case of dust formation occurring early on in a CDS.

The asymmetry of the line profiles point to a pre-existing CSM with a density enhancement inclined 30-45 degrees from edge-on. Combined with the possible detection of a young (3-6 Myr) cluster at the location of the SN and the large inferred $^{56}$Ni mass of 0.22 M$_{\sun}$, it is possible that SN 2011ja had 25-30  M$_{\sun}$ LBV or RSG progenitor that suffered a massive outburst prior to eruption. Although LBVs are generally thought to be the progenitors of Type IIn SNe, recently \citet{2015MNRAS.449.1876S} found that PTF11iqb could have been classified as a normal Type IIP SN had it been discovered just days after the initial classification, as the narrow component quickly disappeared.  This could have been the case with SN 2011ja as well.  If the surrounding CSM was caused by an outburst, it would have likely gone undetected due to the extreme extinction (A$_{v}$$\approx$5.6) in the SN environment.  Deep observations with large ground- or space-based telescopes as SN 2011ja continues to fade will prove essential in putting tighter constraints on the age of the cluster and therefore the mass of SN progenitor.

SN 2011ja seems to share similar properties with the type IIn SNe 1998S and 2005ip, the type Ib/c SN 2006jc, and the type II SNe 2004dj and 2007od.  In all cases interaction with dense CSM creates similar observational signatures, and dust formation at early times in the CDS regardless of initial SN classification.  Particular similarities between SNe 2011ja, 2004dj and 2007od indicate there is a sub-class of Type IIP SNe that have mass-loss histories capable of producing pronounced observational signatures normally reserved for other CCSNe. This only highlights the need for more late time observations of even the most seemingly mundane objects, as there seems to be much we do not know about  pre-supernova mass loss.  

\smallskip\smallskip\smallskip\smallskip
\noindent {\bf ACKNOWLEDGMENTS}
\smallskip
\footnotesize
 This work is based in part on observations from Spitzer Space Telescope and was supported by RSA 1415602 and RSA 1346842,  issued by JPL/Caltech. Based on observations obtained at the Gemini Observatory, which is operated by the Association of Universities for Research in Astronomy, Inc., under a cooperative agreement 
with the NSF on behalf of the Gemini partnership: the National Science Foundation 
(United States), the National Research Council (Canada), CONICYT (Chile), the Australian 
Research Council (Australia), Minist\'{e}rio da Ci\^{e}ncia, Tecnologia e Inova\c{c}\~{a}o 
(Brazil) and Ministerio de Ciencia, Tecnolog\'{i}a e Innovaci\'{o}n Productiva (Argentina). We would like to thank the anonymous referee for their insightful comments during the revision of tihs paper.


\bibliographystyle{mn2e}
\bibliography{SN2011jabib}

\begin{thebibliography}{}

\bibitem[Anderson et al.(2015)]{2015PASA...32...19A} Anderson, J.~P., 
James, P.~A., Habergham, S.~M., Galbany, L., 
\& Kuncarayakti, H.\ 2015, PASA, 32, e019 

\bibitem[\protect\citeauthoryear{{Andrews}, {Gallagher}, {Clayton}, {Sugerman},
  {Chatelain}, {Clem}, {Welch}, {Barlow}, {Ercolano}, {Fabbri}, {Wesson} \&
  {Meixner}}{{Andrews} et~al.}{2010}]{2010ApJ...715..541A}
{Andrews}, J.~E.,  et al.,  2010, ApJ, 715, 541

\bibitem[Andrews et al.(2011a)]{2011ApJ...731...47A} Andrews, J.~E., et al.\ 2011a, ApJ, 731, 47

\bibitem[Andrews et al.(2011b)]{2011AJ....142...45A} Andrews, J.~E., et al.\ 2011b, AJ, 142, 45 

\bibitem[\protect\citeauthoryear{{Andrews}, {Calzetti}, {Chandar}, {Lee},
  {Elmegreen}, {Kennicutt}, {Whitmore}, {Kissel}, {da Silva}, {Krumholz},
  {O'Connell}, {Dopita}, {Frogel} \& {Kim}}{{Andrews}
  et~al.}{2013}]{2013ApJ...767...51A}
{Andrews} J.~E.,  et al.,  2013, ApJ, 767, 51


\bibitem[\protect\citeauthoryear{{Arcavi}, {Gal-Yam}, {Cenko}, {Fox},
  {Leonard}, {Moon}, {Sand}, {Soderberg}, {Kiewe}, {Yaron}, {Becker}, {Scheps},
  {Birenbaum}, {Chamudot} \& {Zhou}}{{Arcavi}
  et~al.}{2012}]{2012ApJ...756L..30A}
{Arcavi} I.,  et al.,
  2012, ApJl, 756, L30
  
  \bibitem[Bevan \& Barlow(2015)]{2015arXiv150900858B} Bevan, A., \& Barlow, M.~J.\ 2015, arXiv:1509.00858

\bibitem[\protect\citeauthoryear{{Barlow}, {Krause}, {Swinyard}, {Sibthorpe},
  {Besel}, {Wesson}, {Ivison}, {Dunne}, {Gear}, {Gomez}, {Hargrave}, {Henning},
  {Leeks}, {Lim}, {Olofsson} \& {Polehampton}}{{Barlow}
  et~al.}{2010}]{2010A&A...518L.138B}
{Barlow} M.~J., et al.,  2010, A\&A, 518, L138

\bibitem[\protect\citeauthoryear{{Cardelli}, {Clayton} \& {Mathis}}{{Cardelli}
  et~al.}{1989}]{1989ApJ...345..245C}
{Cardelli} J.~A.,  {Clayton} G.~C.,    {Mathis} J.~S.,  1989, ApJ, 345, 245

\bibitem[\protect\citeauthoryear{{Chakraborti}, {Ray}, {Smith}, {Ryder},
  {Yadav}, {Sutaria}, {Dwarkadas}, {Chandra}, {Pooley} \& {Roy}}{{Chakraborti}
  et~al.}{2013}]{2013ApJ...774...30C}
{Chakraborti} S., et al., 2013,
  ApJ, 774, 30

\bibitem[\protect\citeauthoryear{{Chevalier}, {Fransson} \&
  {Nymark}}{{Chevalier} et~al.}{2006}]{2006ApJ...641.1029C}
{Chevalier} R.~A.,  {Fransson} C.,    {Nymark} T.~K.,  2006, ApJ, 641, 1029

\bibitem[\protect\citeauthoryear{{Chugai}, {Chevalier} \& {Utrobin}}{{Chugai}
  et~al.}{2007}]{2007ApJ...662.1136C}
{Chugai} N.~N.,  {Chevalier} R.~A.,    {Utrobin} V.~P.,  2007, ApJ, 662, 1136

\bibitem[Draine et al.(2007)]{2007ApJ...663..866D} Draine, B.~T., Dale, 
D.~A., Bendo, G., et al.\ 2007, ApJ, 663, 866 

\bibitem[\protect\citeauthoryear{{Dwek}, {Staguhn}, {Arendt}, {Kovacks}, {Su}
  \& {Benford}}{{Dwek} et~al.}{2014}]{2014ApJ...788L..30D}
{Dwek} E.,  {Staguhn} J.,  {Arendt} R.~G.,  {Kovacks} A.,  {Su} T.,
  {Benford} D.~J.,  2014, ApJl, 788, L30

\bibitem[\protect\citeauthoryear{{Elmhamdi}, {Danziger}, {Chugai},
  {Pastorello}, {Turatto}, {Cappellaro}, {Altavilla}, {Benetti}, {Patat} \&
  {Salvo}}{{Elmhamdi} et~al.}{2003}]{2003MNRAS.338..939E}
{Elmhamdi} A.,  et al., 2003, MNRAS, 338, 939

\bibitem[\protect\citeauthoryear{{Ercolano}, {Barlow} \& {Storey}}{{Ercolano}
  et~al.}{2005}]{2005MNRAS.362.1038E}
{Ercolano} B.,  {Barlow} M.~J.,    {Storey} P.~J.,  2005, MNRAS, 362, 1038

\bibitem[\protect\citeauthoryear{{Ercolano}, {Barlow} \& {Sugerman}}{{Ercolano}
  et~al.}{2007}]{2007MNRAS.375..753E}
{Ercolano} B.,  {Barlow} M.~J.,    {Sugerman} B.~E.~K.,  2007, MNRAS, 375, 753

\bibitem[\protect\citeauthoryear{{Fox}, {Skrutskie}, {Chevalier}, {Kanneganti},
  {Park}, {Wilson}, {Nelson}, {Amirhadji}, {Crump}, {Hoeft}, {Provence},
  {Sargeant}, {Sop}, {Tea}, {Thomas} \& {Woolard}}{{Fox}
  et~al.}{2009}]{2009ApJ...691..650F}
{Fox} O.,  et al., 2009, ApJ, 691, 650

\bibitem[Fox et al.(2010)]{2010ApJ...725.1768F} Fox, O.~D., et al.\ 2010, ApJ, 725, 1768 

\bibitem[\protect\citeauthoryear{{Fransson}, {Chevalier}, {Filippenko},
  {Leibundgut}, {Barth}, {Fesen}, {Kirshner}, {Leonard}, {Li}, {Lundqvist},
  {Sollerman} \& {Van Dyk}}{{Fransson} et~al.}{2002}]{2002ApJ...572..350F}
{Fransson} C. et al., 2002, ApJ, 572, 350

\bibitem[\protect\citeauthoryear{{Fransson}, {Challis}, {Chevalier},
  {Filippenko}, {Kirshner}, {Kozma}, {Leonard}, {Matheson}, {Baron},
  {Garnavich}, {Jha}, {Leibundgut}, {Lundqvist}, {Pun}, {Wang} \&
  {Wheeler}}{{Fransson} et~al.}{2005}]{2005ApJ...622..991F}
{Fransson} C.,  et al.,  2005, ApJ, 622, 991

\bibitem[{{Fransson} {et~al.}(2014){Fransson}, {Ergon}, {Challis}, {Chevalier},
  {France}, {Kirshner}, {Marion}, {Milisavljevic}, {Smith}, {Bufano},
  {Friedman}, {Kangas}, {Larsson}, {Mattila}, {Benetti}, {Chornock}, {Czekala},
  {Soderberg}, \& {Sollerman}}]{2014ApJ...797..118F}
{Fransson} C., et al., 2014, ApJ, 797, 118
  
  \bibitem[Galametz et al.(2011)]{2011A&A...532A..56G} Galametz, M., et al.\ 2011, A\&A, 532, A56 

\bibitem[\protect\citeauthoryear{{Gall}, {Andersen} \& {Hjorth}}{{Gall}
  et~al.}{2011}]{2011A&A...528A..14G}
{Gall} C.,  {Andersen} A.~C.,    {Hjorth} J.,  2011, A\&A, 528, A14

\bibitem[\protect\citeauthoryear{{Gall}, {Hjorth}, {Watson}, {Dwek}, {Maund},
  {Fox}, {Leloudas}, {Malesani} \& {Day-Jones}}{{Gall}
  et~al.}{2014}]{2014Natur.511..326G}{Gall} C., et al.,  2014, Nature, 511, 326

\bibitem[\protect\citeauthoryear{{Gomez}, {Krause}, {Barlow}, {Swinyard},
  {Owen}, {Clark}, {Matsuura}, {Gomez}, {Rho}, {Besel}, {Bouwman}, {Gear},
  {Henning}, {Ivison}, {Polehampton} \& {Sibthorpe}}{{Gomez}
  et~al.}{2012}]{2012ApJ...760...96G}
{Gomez} H.~L., et al., 2012, ApJ, 760, 96

\bibitem[\protect\citeauthoryear{{Hamuy}}{{Hamuy}}{2003}]{2003ApJ...582..905H}
{Hamuy} M.,  2003, ApJ, 582, 905

\bibitem[\protect\citeauthoryear{{Hamuy} \& {Suntzeff}}{{Hamuy} \&
  {Suntzeff}}{1990}]{1990AJ.....99.1146H}
{Hamuy} M.,  {Suntzeff} N.~B.,  1990, AJ, 99, 1146

\bibitem[\protect\citeauthoryear{{Hamuy}, {Suntzeff}, {Gonzalez} \&
  {Martin}}{{Hamuy} et~al.}{1988}]{1988AJ.....95...63H}
{Hamuy} M.,  {Suntzeff} N.~B.,  {Gonzalez} R.,    {Martin} G.,  1988, AJ, 95,
  63

\bibitem[\protect\citeauthoryear{{Hanner}}{{Hanner}}{1988}]{1988ioch.rept...22H}
{Hanner} M.,  1988, Technical report, {Grain optical properties}

\bibitem[{{Hendry} {et~al.}(2005){Hendry}, {Smartt}, {Maund}, {Pastorello},
  {Zampieri}, {Benetti}, {Turatto}, {Cappellaro}, {Meikle}, {Kotak}, {Irwin},
  {Jonker}, {Vermaas}, {Peletier}, {van Woerden}, {Exter}, {Pollacco}, {Leon},
  {Verley}, {Benn}, \& {Pignata}}]{2005MNRAS.359..906H}
{Hendry}, M.~A., {et~al.} 2005, MNRAS, 359, 906

\bibitem[\protect\citeauthoryear{{Hendry}, {Smartt}, {Skillman}, {Evans},
  {Trundle}, {Lennon}, {Crowther} \& {Hunter}}{{Hendry}
  et~al.}{2008}]{2008MNRAS.388.1127H}
{Hendry} M.~A.,  {Smartt} S.~J.,  {Skillman} E.~D.,  {Evans} C.~J.,  {Trundle}
  C.,  {Lennon} D.~J.,  {Crowther} P.~A.,    {Hunter} I.,  2008, MNRAS, 388,
  1127

\bibitem[Immler et al.(2008)]{2008ApJ...674L..85I} Immler, et al.\ 2008, ApJL, 674, L85 

\bibitem[\protect\citeauthoryear{{Kiewe}, {Gal-Yam}, {Arcavi}, {Leonard},
  {Emilio Enriquez}, {Cenko}, {Fox}, {Moon}, {Sand}, {Soderberg} \&
  {CCCP}}{{Kiewe} et~al.}{2012}]{2012ApJ...744...10K}
{Kiewe} M., et al.,   2012, ApJ, 744, 10

\bibitem[\protect\citeauthoryear{{Kotak}, {Meikle}, {Farrah}, {Gerardy},
  {Foley}, {Van Dyk}, {Fransson}, {Lundqvist}, {Sollerman}, {Fesen},
  {Filippenko}, {Mattila}, {Silverman}, {Andersen}, {H{\"o}flich}, {Pozzo} \&
  {Wheeler}}{{Kotak} et~al.}{2009}]{2009ApJ...704..306K}
{Kotak} R.,  et al,  2009, ApJ,
  704, 306

\bibitem[\protect\citeauthoryear{{Kuncarayakti}, {Doi}, {Aldering}, {Arimoto},
  {Maeda}, {Morokuma}, {Pereira}, {Usuda} \& {Hashiba}}{{Kuncarayakti}
  et~al.}{2013}]{2013AJ....146...31K}
{Kuncarayakti} H.,  {Doi} M.,  {Aldering} G.,  {Arimoto} N.,  {Maeda} K.,
  {Morokuma} T.,  {Pereira} R.,  {Usuda} T.,    {Hashiba} Y.,  2013, AJ, 146,
  31

\bibitem[\protect\citeauthoryear{{Larsen}}{{Larsen}}{2004}]{2004A&A...416..537L}
{Larsen} S.~S.,  2004, A\&A, 416, 537

\bibitem[\protect\citeauthoryear{{Lau}, {Herter}, {Morris}, {Li} \&
  {Adams}}{{Lau} et~al.}{2015}]{2015arXiv150307173L}
{Lau} R.~M.,  {Herter} T.~L.,  {Morris} M.~R.,  {Li} Z.,    {Adams} J.~D.,
  2015, arXiv:1503:07173L

\bibitem[\protect\citeauthoryear{{Leitherer}, {Schaerer}, {Goldader},
  {Delgado}, {Robert}, {Kune}, {de Mello}, {Devost} \& {Heckman}}{{Leitherer}
  et~al.}{1999}]{1999ApJS..123....3L}
{Leitherer} C.,  {Schaerer} D.,  {Goldader} J.~D.,  {Delgado} R.~M.~G.,
  {Robert} C.,  {Kune} D.~F.,  {de Mello} D.~F.,  {Devost} D.,    {Heckman}
  T.~M.,  1999, ApJs, 123, 3

\bibitem[\protect\citeauthoryear{{Leonard}, {Filippenko}, {Barth} \&
  {Matheson}}{{Leonard} et~al.}{2000}]{2000ApJ...536..239L}
{Leonard} D.~C.,  {Filippenko} A.~V.,  {Barth} A.~J.,    {Matheson} T.,  2000,
  ApJ, 536, 239

\bibitem[\protect\citeauthoryear{{Levesque}, {Massey}, {Olsen}, {Plez},
  {Josselin}, {Maeder} \& {Meynet}}{{Levesque}
  et~al.}{2005}]{2005ApJ...628..973L}
{Levesque} E.~M.,  {Massey} P.,  {Olsen} K.~A.~G.,  {Plez} B.,  {Josselin} E.,
  {Maeder} A.,    {Meynet} G.,  2005, ApJ, 628, 973

\bibitem[\protect\citeauthoryear{{Lucy}, {Danziger}, {Gouiffes} \&
  {Bouchet}}{{Lucy} et~al.}{1989}]{1989LNP...350..164L}
{Lucy} L.~B.,  {Danziger} I.~J.,  {Gouiffes} C.,    {Bouchet} P.,  1989, in
  {Tenorio-Tagle} G.,  {Moles} M.,   {Melnick} J.,  eds, IAU Colloq. 120:
  Structure and Dynamics of the Interstellar Medium Vol.~350 of Lecture Notes
  in Physics, Berlin Springer Verlag, {Dust Condensation in the Ejecta of SN
  1987 A}.
p.~164

\bibitem[Lyman et al.(2014)]{2014arXiv1406.3667L} Lyman, J., Bersier, D., 
James, P., et al.\ 2014, arXiv:1406.3667 

\bibitem[{{Maeda} {et~al.}(2015){Maeda}, {Hattori}, {Milisavljevic},
  {Folatelli}, {Drout}, {Kuncarayakti}, {Margutti}, {Kamble}, {Soderberg},
  {Tanaka}, {Kawabata}, {Kawabata}, {Yamanaka}, {Nomoto}, {Kim}, {Simon},
  {Phillips}, {Parrent}, {Nakaoka}, {Moriya}, {Suzuki}, {Takaki}, {Ishigaki},
  {Sakon}, {Tajitsu}, \& {Iye}}]{2015arXiv150406668M}
{Maeda}, K., {et~al.} 2015, ArXiv e-prints

\bibitem[\protect\citeauthoryear{{Marconi}, {Oliva}, {van der Werf},
  {Maiolino}, {Schreier}, {Macchetto} \& {Moorwood}}{{Marconi}
  et~al.}{2000}]{2000A&A...357...24M}
{Marconi} A.,  {Oliva} E.,  {van der Werf} P.~P.,  {Maiolino} R.,  {Schreier}
  E.~J.,  {Macchetto} F.,    {Moorwood} A.~F.~M.,  2000, A\&A, 357, 24

\bibitem[\protect\citeauthoryear{{Matheson}, {Filippenko}, {Barth}, {Ho},
  {Leonard}, {Bershady}, {Davis}, {Finley}, {Fisher}, {Gonz{\'a}lez}, {Hawley},
  {Koo}, {Li}, {Lonsdale}, {Schlegel}, {Smith}, {Spinrad} \&
  {Wirth}}{{Matheson} et~al.}{2000}]{2000AJ....120.1487M}
{Matheson} T., et al, 2000, AJ, 120, 1487

\bibitem[\protect\citeauthoryear{{Mathis}, {Rumpl} \& {Nordsieck}}{{Mathis}
  et~al.}{1977}]{1977ApJ...217..425M}
{Mathis} J.~S.,  {Rumpl} W.,    {Nordsieck} K.~H.,  1977, ApJ, 217, 425

\bibitem[{{Matsuura} {et~al.}(2011){Matsuura}, {Dwek}, {Meixner}, {Otsuka},
  {Babler}, {Barlow}, {Roman-Duval}, {Engelbracht}, {Sandstrom},
  {Laki{\'c}evi{\'c}}, {van Loon}, {Sonneborn}, {Clayton}, {Long}, {Lundqvist},
  {Nozawa}, {Gordon}, {Hony}, {Panuzzo}, {Okumura}, {Misselt}, {Montiel}, \&
  {Sauvage}}]{2011Sci...333.1258M}
{Matsuura}, M., {et~al.} 2011, Science, 333, 1258

\bibitem[{{Matsuura} {et~al.}(2015){Matsuura}, {Dwek}, {Barlow}, {Babler},
  {Baes}, {Meixner}, {Cernicharo}, {Clayton}, {Dunne}, {Fransson}, {Fritz},
  {Gear}, {Gomez}, {Groenewegen}, {Indebetouw}, {Ivison}, {Jerkstrand},
  {Lebouteiller}, {Lim}, {Lundqvist}, {Pearson}, {Roman-Duval}, {Royer},
  {Staveley-Smith}, {Swinyard}, {van Hoof}, {van Loon}, {Verstappen}, {Wesson},
  {Zanardo}, {Blommaert}, {Decin}, {Reach}, {Sonneborn}, {Van de Steene}, \&
  {Yates}}]{2015ApJ...800...50M}
{Matsuura}, M., 2015, ApJ, 800, 50

\bibitem[\protect\citeauthoryear{{Mattila}, {Meikle}, {Lundqvist},
  {Pastorello}, {Kotak}, {Eldridge}, {Smartt}, {Adamson}, {Gerardy}, {Rizzi},
  {Stephens} \& {van Dyk}}{{Mattila} et~al.}{2008}]{2008MNRAS.389..141M}
{Mattila} S., et al, 2008, MNRAS, 389, 141

\bibitem[\protect\citeauthoryear{{Mauerhan}, {Williams}, {Smith}, {Smith},
  {Filippenko}, {Hoffman}, {Milne}, {Leonard}, {Clubb}, {Fox} \&
  {Kelly}}{{Mauerhan} et~al.}{2014}]{2014MNRAS.442.1166M}
{Mauerhan} J., et al, 2014, MNRAS, 442, 1166

\bibitem[\protect\citeauthoryear{{Mauerhan}, {Smith}, {Filippenko},
  {Blanchard}, {Blanchard}, {Casper}, {Cenko}, {Clubb}, {Cohen}, {Fuller}, {Li}
  \& {Silverman}}{{Mauerhan} et~al.}{2013}]{2013MNRAS.430.1801M}
{Mauerhan} J.~C., et al.,  2013, MNRAS, 430, 1801

\bibitem[\protect\citeauthoryear{{Mauron} \& {Josselin}}{{Mauron} \&
  {Josselin}}{2011}]{2011A&A...526A.156M}
{Mauron} N.,  {Josselin} E.,  2011, A\&A, 526, A156

\bibitem[\protect\citeauthoryear{{McCray}}{{McCray}}{1993}]{1993ARA&A..31..175M}
{McCray} R.,  1993, ARA\&A, 31, 175

\bibitem[\protect\citeauthoryear{{Meikle}, {Mattila}, {Pastorello}, {Gerardy},
  {Kotak}, {Sollerman}, {Van Dyk}, {Farrah}, {Filippenko}, {H{\"o}flich},
  {Lundqvist}, {Pozzo} \& {Wheeler}}{{Meikle}
  et~al.}{2007}]{2007ApJ...665..608M}
{Meikle} W.~P.~S., et al., 2007,
  ApJ, 665, 608

\bibitem[\protect\citeauthoryear{{Meikle}, {Kotak}, {Farrah}, {Mattila}, {Van
  Dyk}, {Andersen}, {Fesen}, {Filippenko}, {Foley}, {Fransson}, {Gerardy},
  {H{\"o}flich}, {Lundqvist}, {Pozzo}, {Sollerman} \& {Wheeler}}{{Meikle}
  et~al.}{2011}]{2011ApJ...732..109M}
{Meikle} W.~P.~S.,  et al.,  2011, ApJ, 732, 109



\bibitem[\protect\citeauthoryear{{Melena}, {Massey}, {Morrell} \&
  {Zangari}}{{Melena} et~al.}{2008}]{2008AJ....135..878M}
{Melena} N.~W.,  {Massey} P.,  {Morrell} N.~I.,    {Zangari} A.~M.,  2008, AJ,
  135, 878

\bibitem[Micha{\l}owski(2015)]{2015A&A...577A..80M} Micha{\l}owski, M.~J.\ 2015, A$\&$A, 577, A80 

\bibitem[{{Monard} {et~al.}(2011){Monard}, {Milisavljevic}, {Fesen},
  {Pickering}, {Romero-Colmenero}, {Turatto}, {Benetti}, {Pastorello},
  {Valenti}, {Bufano}, {Tomasella}, {Ryder}, {Soderberg}, {Stockdale}, {van
  Dyk}, {Immler}, {Weiler}, \& {Panagia}}]{2011CBET.2946....1M}
{Monard}, L.~A.~G., {et~al.} 2011, Central Bureau Electronic Telegrams, 2946, 1

\bibitem[\protect\citeauthoryear{{Morgan} \& {Edmunds}}{{Morgan} \&
  {Edmunds}}{2003}]{2003MNRAS.343..427M}
{Morgan} H.~L.,  {Edmunds} M.~G.,  2003, MNRAS, 343, 427

\bibitem[\protect\citeauthoryear{{Mouhcine}, {Ferguson}, {Rich}, {Brown} \&
  {Smith}}{{Mouhcine} et~al.}{2005}]{2005ApJ...633..810M}
{Mouhcine} M.,  {Ferguson} H.~C.,  {Rich} R.~M.,  {Brown} T.~M.,    {Smith}
  T.~E.,  2005, ApJ, 633, 810

\bibitem[\protect\citeauthoryear{{Nota}, {Pasquali}, {Clampin}, {Pollacco},
  {Scuderi} \& {Livio}}{{Nota} et~al.}{1996}]{1996ApJ...473..946N}
{Nota} A.,  {Pasquali} A.,  {Clampin} M.,  {Pollacco} D.,  {Scuderi} S.,
  {Livio} M.,  1996, ApJ, 473, 946

\bibitem[\protect\citeauthoryear{{Oey}, {Lamb}, {Kushner}, {Pellegrini} \&
  {Graus}}{{Oey} et~al.}{2013}]{2013ApJ...768...66O}
{Oey} M.~S.,  {Lamb} J.~B.,  {Kushner} C.~T.,  {Pellegrini} E.~W.,    {Graus}
  A.~S.,  2013, ApJ, 768, 66

\bibitem[\protect\citeauthoryear{{Owen} \& {Barlow}}{{Owen} \&
  {Barlow}}{2015}]{2015ApJ...801..141O}
{Owen} P.~J.,  {Barlow} M.~J.,  2015, ApJ, 801, 141

\bibitem[\protect\citeauthoryear{{Poznanski}, {Ganeshalingam}, {Silverman} \&
  {Filippenko}}{{Poznanski} et~al.}{2011}]{2011MNRAS.415L..81P}
{Poznanski} D.,  {Ganeshalingam} M.,  {Silverman} J.~M.,    {Filippenko} A.~V.,
   2011, MNRAS, 415, L81

\bibitem[\protect\citeauthoryear{{Poznanski}, {Prochaska} \&
  {Bloom}}{{Poznanski} et~al.}{2012}]{2012MNRAS.426.1465P}
{Poznanski} D.,  {Prochaska} J.~X.,    {Bloom} J.~S.,  2012, MNRAS, 426, 1465

\bibitem[\protect\citeauthoryear{{Pozzo}, {Meikle}, {Fassia}, {Geballe},
  {Lundqvist}, {Chugai} \& {Sollerman}}{{Pozzo}
  et~al.}{2004}]{2004MNRAS.352..457P}
{Pozzo} M.,  {Meikle} W.~P.~S.,  {Fassia} A.,  {Geballe} T.,  {Lundqvist} P.,
  {Chugai} N.~N.,    {Sollerman} J.,  2004, MNRAS, 352, 457

\bibitem[{{Pozzo} {et~al.}(2006){Pozzo}, {Meikle}, {Rayner}, {Joseph},
  {Filippenko}, {Foley}, {Li}, {Mattila}, \& {Sollerman}}]{2006MNRAS.368.1169P}
{Pozzo}, M., {et~al.} 2006, MNRAS, 368, 1169

\bibitem[\protect\citeauthoryear{{Rossa} \& {Dettmar}}{{Rossa} \&
  {Dettmar}}{2003}]{2003A&A...406..505R}
{Rossa} J.,  {Dettmar} R.-J.,  2003, A\&A, 406, 505

\bibitem[\protect\citeauthoryear{{Sahu}, {Anupama}, {Srividya} \&
  {Muneer}}{{Sahu} et~al.}{2006}]{2006MNRAS.372.1315S}
{Sahu} D.~K.,  {Anupama} G.~C.,  {Srividya} S.,    {Muneer} S.,  2006, MNRAS,
  372, 1315

\bibitem[Seaton(1979)]{1979MNRAS.187P..73S} Seaton, M.~J.\ 1979, MNRAS, 
187, 73P 

\bibitem[\protect\citeauthoryear{{Schlafly} \& {Finkbeiner}}{{Schlafly} \&
  {Finkbeiner}}{2011}]{2011ApJ...737..103S}
{Schlafly} E.~F.,  {Finkbeiner} D.~P.,  2011, ApJ, 737, 103

\bibitem[\protect\citeauthoryear{{Slavin}, {Dwek} \& {Jones}}{{Slavin}
  et~al.}{2015}]{2015ApJ...803....7S}
{Slavin} J.~D.,  {Dwek} E.,    {Jones} A.~P.,  2015, ApJ, 803, 7

\bibitem[\protect\citeauthoryear{{Smartt}}{{Smartt}}{2015}]{2015arXiv150402635S}
{Smartt} S.~J.,  2015, arXiv:1504.02635S

\bibitem[\protect\citeauthoryear{{Smartt}, {Eldridge}, {Crockett} \&
  {Maund}}{{Smartt} et~al.}{2009}]{2009MNRAS.395.1409S}
{Smartt} S.~J.,  {Eldridge} J.~J.,  {Crockett} R.~M.,    {Maund} J.~R.,  2009,
  MNRAS, 395, 1409

\bibitem[\protect\citeauthoryear{{Smith}}{{Smith}}{2007}]{2007AJ....133.1034S}
{Smith} N.,  2007, AJ, 133, 1034

\bibitem[\protect\citeauthoryear{{Smith}, {Arnett}, {Bally}, {Ginsburg} \&
  {Filippenko}}{{Smith} et~al.}{2013}]{2013MNRAS.429.1324S}
{Smith} N.,  {Arnett} W.~D.,  {Bally} J.,  {Ginsburg} A.,    {Filippenko}
  A.~V.,  2013, MNRAS, 429, 1324

\bibitem[\protect\citeauthoryear{{Smith}, {Foley} \& {Filippenko}}{{Smith}
  et~al.}{2008}]{2008ApJ...680..568S}
{Smith} N.,  {Foley} R.~J.,    {Filippenko} A.~V.,  2008, ApJ, 680, 568

\bibitem[\protect\citeauthoryear{{Smith}, {Hinkle} \& {Ryde}}{{Smith}
  et~al.}{2009}]{2009AJ....137.3558S}
{Smith} N.,  {Hinkle} K.~H.,    {Ryde} N.,  2009, AJ, 137, 3558

\bibitem[\protect\citeauthoryear{{Smith}, {Li}, {Miller}, {Silverman},
  {Filippenko}, {Cuillandre}, {Cooper}, {Matheson} \& {Van Dyk}}{{Smith}
  et~al.}{2011}]{2011ApJ...732...63S}
{Smith} N.,  et al., 2011, ApJ, 732, 63

\bibitem[{{Smith} {et~al.}(2015){Smith}, {Mauerhan}, {Cenko}, {Kasliwal},
  {Silverman}, {Filippenko}, {Gal-Yam}, {Clubb}, {Graham}, {Leonard}, {Horst},
  {Williams}, {Andrews}, {Kulkarni}, {Nugent}, {Sullivan}, {Maguire}, {Xu}, \&
  {Ben-Ami}}]{2015MNRAS.449.1876S}
{Smith} N., et al., 2015, MNRAS, 449, 1876


\bibitem[\protect\citeauthoryear{{Smith}, {Mauerhan} \& {Prieto}}{{Smith}
  et~al.}{2014}]{2014MNRAS.438.1191S}
{Smith} N.,  {Mauerhan} J.~C.,    {Prieto} J.~L.,  2014, MNRAS, 438, 1191

\bibitem[\protect\citeauthoryear{{Smith}, {Silverman}, {Chornock},
  {Filippenko}, {Wang}, {Li}, {Ganeshalingam}, {Foley}, {Rex} \&
  {Steele}}{{Smith} et~al.}{2009}]{2009ApJ...695.1334S}
{Smith} N., et al, 2009, ApJ, 695, 1334

\bibitem[\protect\citeauthoryear{{Smith}, {Silverman}, {Filippenko}, {Cooper},
  {Matheson}, {Bian}, {Weiner} \& {Comerford}}{{Smith}
  et~al.}{2012}]{2012AJ....143...17S}
{Smith} N.,  {Silverman} J.~M.,  {Filippenko} A.~V.,  {Cooper} M.~C.,
  {Matheson} T.,  {Bian} F.,  {Weiner} B.~J.,    {Comerford} J.~M.,  2012, AJ,
  143, 17
  
  \bibitem[Stanghellini et al.(2015)]{2015arXiv150802754S} Stanghellini, L., 
Magrini, L., \& Casasola, V.\ 2015, arXiv:1508.02754 

\bibitem[\protect\citeauthoryear{{Sugerman}, {Andrews}, {Barlow}, {Clayton},
  {Ercolano}, {Ghavamian}, {Kennicutt} Jr., {Krause}, {Meixner} \&
  {Otsuka}}{{Sugerman} et~al.}{2012}]{2012ApJ...749..170S}
{Sugerman} B.~E.~K., et al., 2012, ApJ, 749, 170

\bibitem[\protect\citeauthoryear{{Sugerman}, {Ercolano}, {Barlow}, {Tielens},
  {Clayton}, {Zijlstra}, {Meixner}, {Speck}, {Gledhill}, {Panagia}, {Cohen},
  {Gordon}, {Meyer}, {Fabbri}, {Bowey}, {Welch}, {Regan} \&
  {Kennicutt}}{{Sugerman} et~al.}{2006}]{2006Sci...313..196S}
{Sugerman} B.~E.~K., et al., 2006, Science, 313, 196

\bibitem[\protect\citeauthoryear{{Suntzeff}, {Hamuy}, {Martin}, {Gomez} \&
  {Gonzalez}}{{Suntzeff} et~al.}{1988}]{1988AJ.....96.1864S}
{Suntzeff} N.~B.,  {Hamuy} M.,  {Martin} G.,  {Gomez} A.,    {Gonzalez} R.,
  1988, AJ, 96, 1864

\bibitem[{{Taubenberger} {et~al.}(2011){Taubenberger}, {Navasardyan}, {Maurer},
  {Zampieri}, {Chugai}, {Benetti}, {Agnoletto}, {Bufano}, {Elias-Rosa},
  {Turatto}, {Patat}, {Cappellaro}, {Mazzali}, {Iijima}, {Valenti},
  {Harutyunyan}, {Claudi}, \& {Dolci}}]{2011MNRAS.413.2140T}
{Taubenberger}, S., {et~al.} 2011, MNRAS, 413, 2140


\bibitem[\protect\citeauthoryear{{Temim}, {Dwek}, {Tchernyshyov}, {Boyer},
  {Meixner}, {Gall} \& {Roman-Duval}}{{Temim}
  et~al.}{2015}]{2015ApJ...799..158T}
{Temim} T.,  {Dwek} E.,  {Tchernyshyov} K.,  {Boyer} M.~L.,  {Meixner} M.,
  {Gall} C.,    {Roman-Duval} J.,  2015, ApJ, 799, 158

\bibitem[\protect\citeauthoryear{{Vink{\'o}}, {S{\'a}rneczky}, {Balog},
  {Immler}, {Sugerman}, {Brown}, {Misselt}, {Szab{\'o}}, {Csizmadia}, {Kun},
  {Klagyivik}, {Foley}, {Filippenko}, {Cs{\'a}k} \& {Kiss}}{{Vink{\'o}}
  et~al.}{2009}]{2009ApJ...695..619V}
{Vink{\'o}} J., et al. 2009, ApJ, 695, 619

\bibitem[{{Vink{\'o}} {et~al.}(2006){Vink{\'o}}, {Tak{\'a}ts}, {S{\'a}rneczky},
  {Szab{\'o}}, {M{\'e}sz{\'a}ros}, {Csorv{\'a}si}, {Szalai}, {G{\'a}sp{\'a}r},
  {P{\'a}l}, {Csizmadia}, {K{\'o}sp{\'a}l}, {R{\'a}cz}, {Kun}, {Cs{\'a}k},
  {F{\"u}r{\'e}sz}, {DeBond}, {Grunhut}, {Thomson}, {Mochnacki}, \&
  {Koktay}}]{2006MNRAS.369.1780V}
{Vink{\'o}}, J., {et~al.} 2006, MNRAS, 369, 1780

\bibitem[\protect\citeauthoryear{{Welch}, {Clayton}, {Campbell}, {Barlow},
  {Sugerman}, {Meixner} \& {Bank}}{{Welch} et~al.}{2007}]{2007ApJ...669..525W}
{Welch} D.~L.,  {Clayton} G.~C.,  {Campbell} A.,  {Barlow} M.~J.,  {Sugerman}
  B.~E.~K.,  {Meixner} M.,    {Bank} S.~H.~R.,  2007, ApJ, 669, 525

\bibitem[Wesson et al.(2015)]{2015MNRAS.446.2089W} Wesson, R., Barlow, 
M.~J., Matsuura, M., \& Ercolano, B.\ 2015, MNRAS, 446, 2089 

\bibitem[\protect\citeauthoryear{{Wooden}, {Rank}, {Bregman}, {Witteborn},
  {Tielens}, {Cohen}, {Pinto} \& {Axelrod}}{{Wooden}
  et~al.}{1993}]{1993ApJS...88..477W}
{Wooden} D.~H.,  {Rank} D.~M.,  {Bregman} J.~D.,  {Witteborn} F.~C.,  {Tielens}
  A.~G.~G.~M.,  {Cohen} M.,  {Pinto} P.~A.,    {Axelrod} T.~S.,  1993, ApJs,
  88, 477

\end{thebibliography}

\end{document}